\documentstyle[sprocl,epsf,rotating,psfig]{article}

\newcommand{\lsim}{\raisebox{-0.13cm}{~\shortstack{$<$ \\[-0.07cm] $\sim$}}~}
\newcommand{\gsim}{\raisebox{-0.13cm}{~\shortstack{$>$ \\[-0.07cm] $\sim$}}~}

\newcommand{\ra}{\rightarrow}
\newcommand{\ee}{e^+e^-}
\newcommand{\s}{\\ \vspace*{-3.5mm} }
\newcommand{\nn}{\noindent}
\newcommand{\non}{\nonumber}
\newcommand{\beq}{\begin{eqnarray}}
\newcommand{\eeq}{\end{eqnarray}}

\newcommand{\MSSM}{MSSM}

\newcommand{\tb}{\mbox{tg}\beta}

\newcommand{\ctg}{\mbox{ctg}}

\newcommand{\GeV}{\rm GeV}

\begin{document}

\thispagestyle{empty}

\begin{flushright}
PM--97/51 \\
December 1997
\end{flushright}

\vspace*{1cm}

\begin{center}

\Large{\bf Decays of the Higgs Bosons} 

\vspace*{1cm} 

\large{\sc Abdelhak DJOUADI} 

\vspace*{1cm}

Laboratoire de Physique Math\'ematique et Th\'eorique, UMR--CNRS, \\
Universit\'e Montpellier II, F--34095 Montpellier Cedex 5, France. \\
E-mail: djouadi@lpm.univ-montp2.fr \\

\vspace*{2cm} 

{\bf ABSTRACT}

\bigskip 

\end{center} 

\noindent We review the decay modes of the Standard Model Higgs boson and 
those of the neutral and charged Higgs particles of the Minimal Supersymmetric 
extension of the Standard Model. Special emphasis will be put on higher--order 
effects.

\vspace*{3cm}

\begin{center}
Talk given at the {\it International Workshop on Quantum Effects in the MSSM}
\\
Barcelona, Spain, September 9--13, 1997. 
\end{center}

\setcounter{page}{0}

\newpage

\title{Decays of the Higgs Bosons} 
\author{Abdelhak DJOUADI} 
\address{Laboratoire de Physique Math\'ematique et Th\'eorique, UMR--CNRS, \\
Universit\'e Montpellier II, F--34095 Montpellier Cedex 5, France. \\
E-mail: djouadi@lpm.univ-montp2.fr}
\maketitle\abstracts{
We review the decay modes of the Standard Model Higgs boson and those of the 
neutral and charged Higgs particles of the Minimal Supersymmetric extension of 
the Standard Model. Special emphasis will be put on higher--order effects.}

\section{Introduction}
The experimental observation of scalar Higgs particles is crucial for 
our present understanding of the mechanism of electroweak symmetry breaking.
Thus the search for Higgs bosons is one of the main entries in the
LEP2 agenda, and will be one of the major goals of future colliders
such as the Large Hadron Collider LHC and the future Linear $\ee$ Collider
LC. Once the Higgs boson is found, it will be of utmost importance to
perform a detailed investigation of its fundamental properties, a crucial
requirement to establish the Higgs mechanism as the basic way to
generate the masses of the known particles. To this end, a very precise
prediction of the production cross sections and of the branching ratios
for the main decay channels is mandatory. \s 

In the Standard Model (SM), one doublet of scalar fields is needed for
the electroweak symmetry breaking, leading to the existence of one
neutral scalar particle \cite{HHG} $H^0$. Once $M_{H^0}$ is fixed, the
profile of the Higgs boson is uniquely determined at tree level: the
couplings to fermions and gauge bosons are set by their masses and all
production cross sections, decay widths and branching ratios can be
calculated unambiguously \cite{REV}. Unfortunately, $M_{H^0}$ is a
free parameter.  {}From the direct search at LEP1 and LEP2 we know that it
should be larger than \cite{LEPbound} 77.1 GeV. Triviality restricts 
the Higgs particle to be lighter than about 1 TeV; theoretical arguments based
on Grand Unification at a scale $\sim10^{16}$ GeV suggest however,
that the preferred mass region will be 100 GeV $ \lsim M_{H^0} \lsim$
200 GeV; for a recent summary, see Ref.~\cite{DATA}.  \s

In supersymmetric (SUSY) theories, the Higgs sector is extended to contain at
least two isodoublets of scalar fields. In the Minimal Supersymmetric Standard
Model (MSSM) this leads to the existence of five physical Higgs 
particles \cite{HHG}: two
CP-even Higgs bosons $h$  and $H$, one CP-odd or pseudoscalar Higgs boson $A$,
and two charged Higgs particles $H^\pm$. Besides the four masses,
two additional parameters are needed: the ratio of the two vacuum expectation
values, $\tb$, and a mixing angle $\alpha$ in the CP-even sector. However, only
two of these parameters are independent: choosing the pseudoscalar mass $M_A$
and $\tb$ as inputs, the structure of the MSSM Higgs sector is entirely
determined at lowest order. However, large SUSY radiative corrections 
\cite{RC,quiros} affect the Higgs masses and couplings, introducing new [soft
SUSY-breaking] parameters in the Higgs sector. If in addition relatively light
genuine supersymmetric particles are allowed, the whole set of SUSY parameters
will be needed to describe the MSSM Higgs boson properties unambiguously. \s 

In this talk, I will discuss the decay widths and branching ratios of the Higgs 
bosons in the SM and in the MSSM. Special emphasis will be put on higher--order
effects such as QCD and electroweak corrections, three--body decay modes and 
SUSY--loop contributions. For details on the MSSM Higgs boson masses and 
couplings including radiative corrections \cite{RC}, and in general on the 
parameters of the MSSM, we refer the reader to \cite{HHG} or to the 
reviews in Refs.~\cite{DATA,quiros,habil,ho}. 

\section{Decay Modes in the Standard Model}

\subsection{Decays to quarks and leptons} 

The partial widths for decays to massless quarks directly coupled to the SM 
Higgs particle, including the ${\cal O}(\alpha_{s}^{2})$ radiative corrections 
\cite{QCD}, is given by \cite{drees,russ}
\begin{eqnarray}
\Gamma [H^0  \ra Q{\overline{Q}}] = \frac{3 G_F M_{H^0} } {4\sqrt{2}\pi}
\overline{m}_Q^2(M_{H^0} ) \left[ 1 + 5.67 \frac{\alpha_s}
{\pi} + (35.94 - 1.36 N_F) \frac{\alpha_s^2}{\pi^2} \right]
\end{eqnarray}
in the ${\overline{\rm MS}}$ renormalization scheme. The ${\cal O}( 
\alpha_{s}^{3})$ QCD radiative corrections are also known \cite{russ}. 
Large logarithms are
resummed by using the running quark mass $\overline{m}_{Q}(M_{H^0})$ and
the strong coupling $\alpha_s(M_{H^0})$ both defined at the scale $M_{H^0}$.
The quark masses can be neglected in the
phase space and in the matrix element except for decays in the
threshold region, where the next-to-leading-order QCD corrections are 
given in terms of the quark {\it pole} mass $M_Q$ \cite{drees}. \s

The relation between the perturbative {\it pole} quark mass ($M_Q$)
and the running $\overline{\rm MS}$ mass (${\overline{m}}_{Q}$) at the
scale of the pole mass can be expressed as \cite{broad}
\begin{equation} \label{run-pole}
{\overline{m}}_{Q}(M_{Q})= M_{Q} [ 1+ 4\alpha_{s}(M_Q)/(3\pi) + K_Q 
( \alpha_s(M_Q)/ \pi)^2 ]^{-1}
\end{equation}
where the numerical values of the NNLO coefficients are given by
$K_t\sim 10.9$, $K_b \sim 12.4$ and $K_c \sim 13.4$.  Since the
relation between the pole mass $M_{c}$ of the charm quark and the
${\overline{\rm MS}}$ mass ${\overline{m}}_{c}(M_{c})$ evaluated at
the pole mass is badly convergent \cite{broad}, the running quark
masses ${\overline{m}}_{Q}(M_{Q})$ are adopted as starting points, because
these are directly determined from QCD spectral sum rules \cite{narison} for
the $b$ and $c$ quarks. The input pole mass values and corresponding running 
masses are presented in Table 1 for charm and bottom quarks. In the case of
the top quark, with $\alpha_s=118$ and $M^{\rm pt2}_t=175$ GeV, one has
$\overline{m}_Q (M_t)=167.4$ GeV and $M_t=177.1$ GeV. 

\vspace*{-4mm}

\begin{table}[htbp]
\caption[]{Quark mass values for the $\overline{\rm MS}$ mass and 
  the two different definitions of the pole masses. 
  $\alpha_s(M_Z)=0.118$ and the bottom and charm mass
  values are taken from Ref.~\cite{narison}.}
\begin{center}
\begin{tabular}{|c|c|cc|c|} \hline
&$ \alpha_{s}(M_{Z}) $
& $ {\overline{m}}_{Q}(M_{Q})$ &  $M_{Q} = M_{Q}^{\rm pt2} $
& ${\overline{m}}_{Q} (\mu = 100~\GeV) $ \\ \hline
$b$ & $0.112$ & $(4.26 \pm 0.02)~\GeV$ & $(4.62 \pm 0.02)~\GeV$
& $(3.04 \pm 0.02)~\GeV$ \\
    & $0.118$ & $(4.23 \pm 0.02)~\GeV$ & $(4.62 \pm 0.02)~\GeV$
& $(2.92 \pm 0.02)~\GeV$ \\
    & $0.124$ & $(4.19 \pm 0.02)~\GeV$ & $(4.62 \pm 0.02)~\GeV$
& $(2.80 \pm 0.02)~\GeV$ \\ \hline
$c$ & $0.112$ & $(1.25 \pm 0.03)~\GeV$ & $(1.42 \pm 0.03)~\GeV$
& $(0.69 \pm 0.02)~\GeV$ \\
    & $0.118$ & $(1.23 \pm 0.03)~\GeV$ & $(1.42 \pm 0.03)~\GeV$
& $(0.62 \pm 0.02)~\GeV$ \\
    & $0.124$ & $(1.19 \pm 0.03)~\GeV$ & $(1.42 \pm 0.03)~\GeV$
& $(0.53 \pm 0.02)~\GeV$ \\ \hline
\end{tabular}
\end{center}
\end{table} 

\noindent The evolution from $M_{Q}$ upwards to a renormalization scale $\mu$ is
given by \cite{runmass}
\begin{eqnarray}
&& \hspace*{1.5cm}{\overline{m}}_{Q}\,(\mu )={\overline{m}}_{Q}\,(M_{Q})
\,\frac{c\,[\alpha_{s}\,(\mu)/\pi ]}{c\, [\alpha_{s}\,(M_{Q})/\pi ]}
\label{eq:msbarev} \\
c(x)&=&\left(9x/2 \right)^{\frac{4}{9}} \, [1+0.895x+1.371\,x^{2}]
\hspace{1.35cm} \mbox{for} \hspace{.2cm} M_{s}\,<\mu\,<M_{c} \non \\
c(x)&=&\left(25x/6 \right)^{\frac{12}{25}} \, [1+1.014x+1.389\,x^{2}]
\hspace{1.0cm} \mbox{for} \hspace{.2cm} M_{c}\,<\mu\,<M_{b} \non \\
c(x)&=&\left(23x/6\right)^{\frac{12}{23}} \, [1+1.175x+1.501\,x^{2}]
\hspace{1cm} \mbox{for} \hspace{.2cm} M_{b}\,<\mu \,< M_t \non \\
c(x)&=&\left(7x/2 \right)^{\frac{4}{7}} \, [1+1.398x+1.793\,x^{2}]
\hspace{1.35cm} \mbox{for} \hspace{.2cm} M_{t}\,<\mu \non 
\end{eqnarray}
For the charm quark mass the evolution is determined by
eq.~(\ref{eq:msbarev}) up to the scale $\mu=M_b$, while for scales
above the bottom mass the evolution must be restarted at $M_Q = M_b$.
The values of the running $b,c$ masses at the scale $\mu \sim M_H =100$ GeV
are typically 35\% (60\%) smaller
than the bottom (charm) pole masses $M_b^{\rm pt2}$ ($M_c^{\rm pt2})$. \s

The Higgs boson decay width into leptons is obtained by dividing eq.~(1) 
by the color factor $N_c=3$ and by switching off the QCD corrections. 
In the case of the $t\bar t$ decays of the standard Higgs boson, the
${\cal O} (\alpha_s)$ QCD corrections are known exactly \cite{drees}.  
The ${\cal O} (\alpha_s^2)$ QCD corrections have been computed recently 
in Ref.~\cite{mathias1}: compared to the Born term, they are of the order 
of a few percent in the on--shell scheme, but in the ${\rm \overline{MS}}$
scheme, they are very small and can be neglected.  Note that the 
below-threshold (three-body) decays $H\to t\bar t^* \to
t\bar b W^-$ into off-shell top quarks may be sizeable \cite{1OFF} and
should be taken into account for Higgs boson masses close to threshold. \s

Finally, the electroweak corrections to heavy quark and lepton decays in the
intermediate Higgs mass range are small \cite{rcelw} and could thus
be neglected. For large Higgs masses the electroweak corrections due
to the enhanced self-coupling of the Higgs bosons are also quite small 
\cite{rcelw}.

\subsection{Decays to gluons and electroweak gauge bosons} 

The decay of the Higgs boson to gluons is mediated by heavy quark
loops in the SM; the partial width in lowest order is given in \cite{hgglo}.  
QCD radiative corrections \cite{higgsqcd,hgg} are built up by the exchange 
of virtual gluons,
gluon radiation from the quark loop and the splitting of a
gluon into unresolved two gluons and $N_F$  quark-antiquark pair. The 
partial decay width, in the limit $m_t \gg M_H$ which is a good approximation,
and including NLO QCD corrections, is given by
\begin{eqnarray}
\Gamma [ H^0 \ra gg] = \frac{G_{F} \alpha_{s}^2 M_{H^0}^{3}} {36  
\sqrt{2} \pi^{3}}
\left[ 1+ \frac{\alpha_s}{\pi} \left( \frac{95}{4}
-\frac{7}{6}N_{F} +\frac{33-2N_{F}} {6}\log \frac{\mu ^{2}}{M_{H^0}^{2}}
 \right) \right]
\end{eqnarray}
Here $\mu \sim M_{H^0}$ and $\alpha_s\equiv \alpha_s^{N_F}(\mu^2)$. 
The radiative corrections are very large, nearly doubling the partial
width.  Since $b$ quarks, and eventually $c$ quarks, can in principle
be tagged experimentally, it is physically meaningful to include gluon
splitting $g^*\,\ra\, b{\overline{b}} \; (c{\overline{c}})$ in
$H^0\ra\,gg^*\ra\, gb{\overline{b}} \, (c{\overline{c}})$ decays to the
inclusive decay probabilities $\Gamma(H^0 \ra b\bar{b}+ \dots)$ {\it
etc.} \cite{QCD}. The contribution of $b,c$ quark final states to the
coefficient in front of $\alpha_s$ in eq.~(4) is: $-7/3+[\log M_{H^0}^{2}/
M_{b}^{2} +\log M_{H^0}^{2}/M_{c}^{2}]/3$. 
Separating this contribution generates large
logarithms, which can be effectively absorbed by defining the number
of active flavors in the gluonic decay mode. The contributions of the
subtracted flavors will be then added to the corresponding
heavy quark decay modes. \s

Since the two--loop QCD corrections to the $H^0 \ra gg$ decay mode
turn out to be large, one may wonder whether the perturbation series is 
in danger. However, recently the three--loop QCD corrections to this decay
have been calculated \cite{mathias2} in the infinitely heavy quark limit,
$m_t \gg M_{H}^0$. The correction for $N_F=5$ if of order 20\% of the Born 
term and 30\% of the NLO term, therefore showing a good convergence behavior 
of the perturbative series. \s

The decays of the Higgs boson to $\gamma \gamma$ and $\gamma Z$ \cite{hgglo}, 
mediated by $W$ and heavy fermion loops are very rare  with
branching ratios of ${\cal O}(10^{-3})$. However, they are interesting  since
they provide a way to count the number of heavy particles which couple to the
Higgs bosons, even if they are too heavy to be produced directly.  Indeed, 
since the couplings of the loop particles  are proportional to their masses,
they balance the decrease of the triangle  amplitude with increasing mass, and
the particles do not decouple for large masses. QCD radiative corrections to
the quark loops are rather small \cite{higgsqcd} and can be neglected. \s

Finally, above the $WW$ and $ZZ$ decay thresholds, the decay of the Higgs
boson into pairs of massive gauge bosons \cite{hvv0} [$\delta_{W}(\delta_Z) 
=2(1)$]
\begin{eqnarray}
\Gamma [H^0 \ra VV] = \frac{G_F M_{H^0}^3}{16 \sqrt{2} \pi} \, \delta_V \,
\sqrt{1-4x} \, (1-4x +12x^2) \ \ , \ \ x= \frac{M_V^2}{M_{H^0}^2} 
\end{eqnarray}
becomes the dominant mode.
Electroweak corrections are small in the intermediate mass range \cite{rcelw} 
and thus can be neglected. Higher order
corrections due to the self-couplings of the Higgs particles are
sizeable \cite{hvvlam} for $M_{H^0} \gsim 400$ GeV and should be taken into
account. Below the $WW/ZZ$ threshold, the decay modes into off-shell gauge
bosons are important. For instance, for $M_{H^0} \gsim 130$ GeV, the Higgs 
boson decay into 
$WW$ with one off--shell $W$ boson \cite{hvv} starts to dominate over the 
$H^0 \ra b\bar{b}$ mode. In fact even Higgs decays into two off--shell 
gauge bosons \cite{2OFF} can be important. The branching ratios for the latter 
reach the percent level for Higgs masses above about 100 (110) GeV for both
$W(Z)$ boson pairs off-shell. For higher masses, it is sufficient to
allow for one off-shell gauge boson only. The decay width can be cast into
the form \cite{2OFF}: 
\begin{small}
\begin{eqnarray}
\Gamma(H^0\ra V^*V^*) = \frac{1}{\pi^2}\int_0^{M_{H^0}^2} \hspace*{-0.4cm} 
\frac{dq_1^2 M_V \Gamma_V}{(q_1^2 - M_V^2)^2 + M_V^2 \Gamma_V^2}
\int_0^{(M_{H^0}-Q_1)^2} \hspace*{-1cm} \frac{dq_2^2 M_V \Gamma_V} {(q_2^2 
- M_V^2)^2 + M_V^2 \Gamma_V^2} \Gamma_0
\end{eqnarray}
\end{small}
with $q_1^2, q_2^2$ being the squared invariant masses of the virtual 
bosons, $M_V$ and $\Gamma_V$ their masses and total decay widths, and with
$\lambda(x,y;z)=(1-x/z-y/z)^2-4xy/z^2$, $\delta'_W = 1$, $\delta'_Z = 7/12 
- 10\sin^2\theta_W/9 + 40 \sin^4\theta_W/27$, $\Gamma_0$ is 
\begin{eqnarray}
\Gamma_0 = \delta_V' \frac{G_F M_{H^0}^3}{16\sqrt{2}\pi}
\sqrt{\lambda(q_1^2,q_2^2;M_{H^0}^2)} \left[ \lambda(q_1^2,q_2^2;M_{H^0}^2) +
12 q_1^2q_2^2/ M_{H^0}^4 \right] 
\end{eqnarray}

\subsection{Total Decay Width and Branching Ratios}

The total decay width and the branching ratios of the SM Higgs boson 
are shown in Fig.~1. In the ``low mass" range, $M_{H^0}\lsim 140$ GeV, 
the main decay mode is by far $ H^0 \ra b\bar{b}$ with ${\rm BR} \sim 
90\%$ followed by the decays into $c\bar{c}$ and $\tau^+\tau^-$ with 
${\rm BR} \sim 5\%$. Also of significance, the $gg$ decay with ${\rm BR} 
\sim 5\%$ for $M_{H^0} \sim 120$ GeV. The  $\gamma\gamma$ and $Z \gamma$ 
decays are rare, ${\rm BR} \sim {\cal O }(10^{-3})$. In the ``high mass" 
range, $M_{H^0}\gsim 140$ GeV, the Higgs bosons decay into $WW$ and $ZZ$
pairs, with one virtual gauge boson below the threshold. For $M_{H^0} \gsim
2M_Z$, it decays exclusively into these channels with a BR of 2/3 for 
$WW$ and 1/3 for $ZZ$. The opening of the $t\bar{t}$ channel does not alter 
significantly this pattern. \s

In the low mass range, the Higgs boson is very narrow $\Gamma_{H^0} <10$ MeV,
but the width becomes rapidly wider for masses larger than 130 GeV,
reaching $\sim 1$ GeV at the $ZZ$ threshold; the Higgs decay width cannot be
measured directly [at the LHC or at an $e^+e^-$ LC] in the mass range below 
250 GeV. For large masses, 
$M_{H^0} \gsim 500$ GeV, the Higgs boson becomes obese: its decay width becomes 
comparable to its mass. \s

\begin{figure}[hbtp]
\begin{center}
\vspace*{-.1cm}

\hspace*{-1.5cm}
\begin{turn}{-90}%
\epsfxsize=6.cm \epsfbox{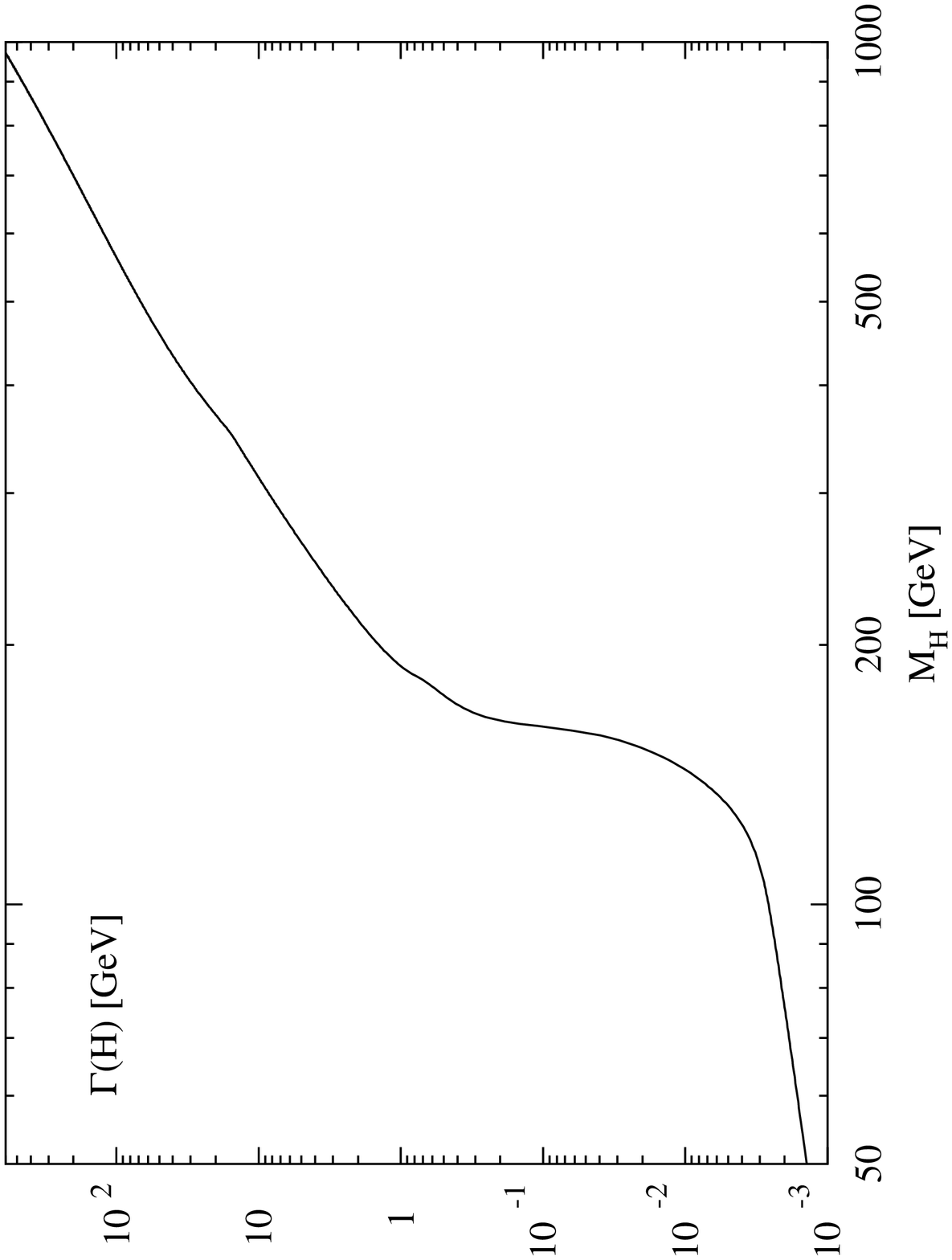}
\end{turn}
\vspace*{.3cm}

\hspace*{-1.5cm}
\begin{turn}{-90}%
\epsfxsize=6.cm \epsfbox{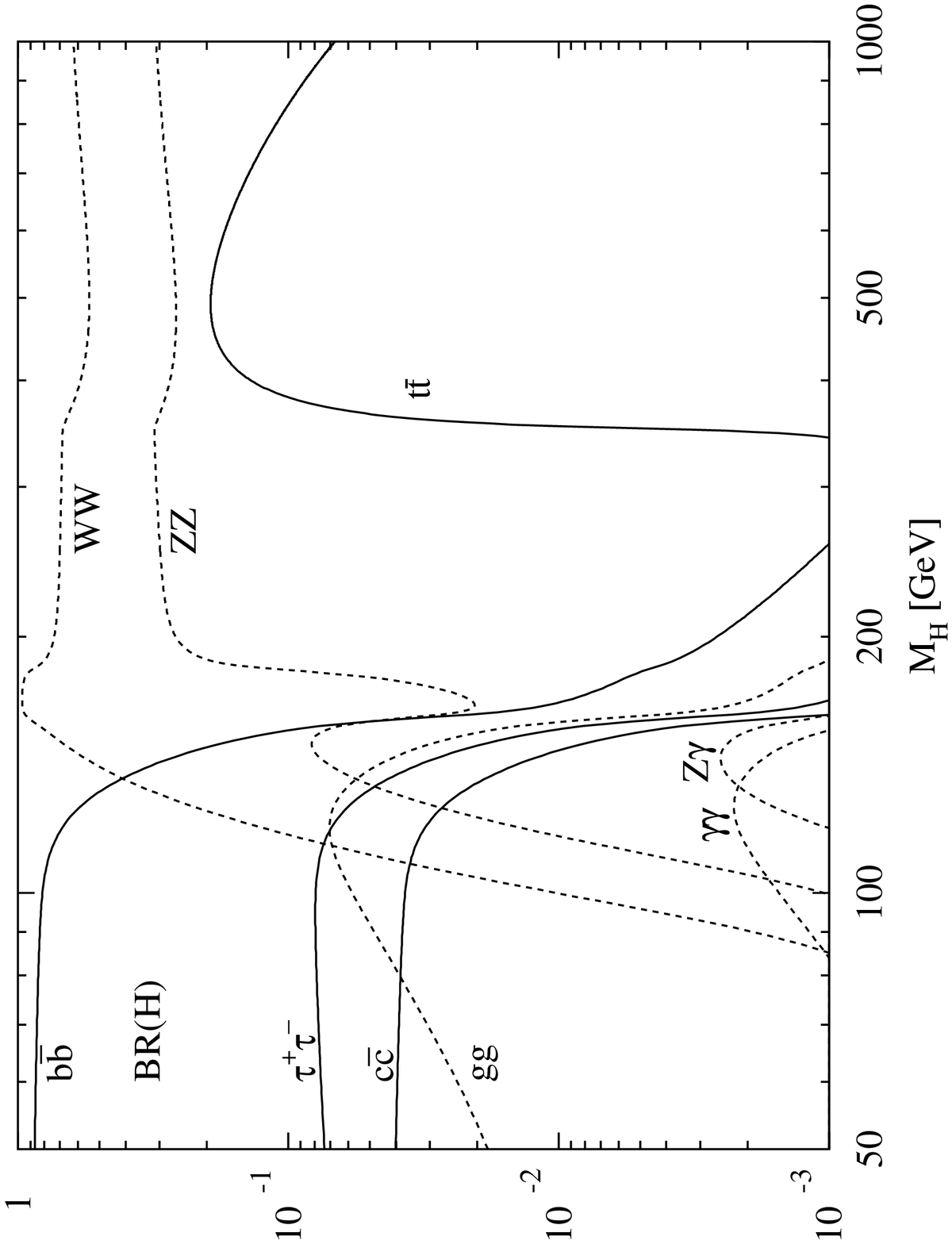}
\end{turn}
\vspace*{-0.2cm}
\end{center}
\caption[]{\it Total decay width $\Gamma(H^0)$ in GeV and the main branching
ratios $BR(H^0)$ of the Standard Model Higgs decay channels.} 
\end{figure}

\section{MSSM Higgs Sector: Standard Decays and Corrections}

\subsection{Higgs boson masses and couplings} 
 
In the MSSM, the Higgs sector \cite{HHG} is highly constrained since there 
are only two free parameters at tree--level: a Higgs mass parameter 
[generally $M_A$] and the ratio of the two vacuum expectation values 
$\tb$ [which in SUSY--GUT models with Yukawa coupling unification is 
forced to be either small, $\tb \sim 1.5$, or large, $\tb \sim$ 30--50]. 
The radiative corrections in the Higgs 
sector \cite{RC}  change significantly the relations between the Higgs boson
masses and couplings and shift the mass of the lightest CP--even Higgs boson 
upwards. The leading part of this correction grows as the fourth power of 
the top quark mass and logarithmically with the common squark mass, 
and can be parameterized by:
$\epsilon = 3 G_{F} m_t^4 /(\sqrt{2}\pi^2 \sin^2\beta) \times
\log (1+M_{\tilde{q}}^2/m_t^2)$. 
The CP--even [and the charged] Higgs boson masses are then given in terms 
of $M_A$, $\tb$ and the parameter $\epsilon$ as [$a=M_A^2$ and $z=M_Z^2$ for 
short]
\begin{eqnarray}
M_{h,H}^2 = \frac{1}{2} [ a+z+\epsilon \mp \sqrt{(a+z+\epsilon)^2- 4
az\cos^2 2\beta - 4\epsilon (a\sin^2\beta + z \cos^2\beta)} ] \non \\ 
M_{H^\pm}^2 = M_A^2 +M_W^2 \hspace*{4cm}
\end{eqnarray}
The decay pattern of the MSSM Higgs bosons is determined to a large extent 
by their couplings to fermions and gauge bosons, which in general depend 
strongly on $\tb$ and the mixing angle $\alpha$ in the CP--even sector, 
which reads 
\begin{eqnarray}
\tan 2\alpha = \tan 2\beta \  (a+z)/(a-z+\epsilon/\cos 2\beta) \ , 
\hspace*{1cm} - \pi/2 \leq \alpha \leq 0 
\end{eqnarray}
The pseudoscalar and charged Higgs boson couplings to down (up) type fermions 
are (inversely) proportional to $\tb$; the pseudoscalar $A$ has no tree level 
couplings to gauge bosons.  For the CP--even Higgs bosons, the couplings to 
down (up) type fermions are enhanced (suppressed) compared to the SM Higgs 
couplings [$\tb>1$]; the couplings to gauge bosons are suppressed by 
$\sin/\cos(\beta-\alpha)$ factors; see Table 2. Note also that the couplings
of the $h$ and $H$ bosons to $ZA$ and $W^+H^-$ pairs are proportional  
$\cos$ and $\sin(\beta-\alpha)$ respectively, while the 
$W^+H^-A$ coupling is not suppressed by these factors.  

\smallskip 

\nn {\small Table 2: Higgs couplings to fermions and gauge bosons normalized
to the SM Higgs couplings, and their limit for $M_A \gg M_Z$.} 
\begin{center}
\begin{tabular}{|c|c|c|c|c|} \hline
$\ \ \ \Phi \ \ \ $ &$ g_{\Phi \bar{u}u} $      & $ g_{\Phi \bar{d} d} $ &
$g_{ \Phi VV} $ \\ \hline
$h$  & \ $\; \cos\alpha/\sin\beta \rightarrow 1     \; $ \ & \ $ \; -\sin\alpha/
\cos\beta \rightarrow 1 \; $ \ & \ $ \; \sin(\beta-\alpha) \rightarrow 1 \; 
$ \ \\
 $H$  & \   $\; \sin\alpha/\sin\beta \rightarrow 1/\tb \; $ \ & \ $ \; 
\cos\alpha/ \cos\beta \rightarrow \tb \; $ \ & \ $ \; \cos(\beta-\alpha) 
\rightarrow 0 \; $ 
\ \\ $A$  & \ $\; 1/ \tb \; $\ & \ $ \; \tb \; $ \   & \ $ \; 0 \; $ \ \\ \hline
\end{tabular}
\end{center}
\bigskip 

\newpage 

\subsection{Decays to quarks and leptons} 

The partial decay widths of the MSSM CP--even neutral Higgs bosons $h$ and 
$H$ to fermions are the same in the SM case with properly the
modified Higgs boson couplings defined in Tab.~2. For massless quarks, the QCD
corrections for scalar, pseudoscalar and charged Higgs boson decays
are similar  to the SM case \cite{drees,russ}, $i.e.$ the Yukawa and QCD 
couplings are evaluated at the scale of the Higgs boson mass. \s

In the threshold regions, mass effects play a significant role, in
particular for the pseudoscalar Higgs boson, which has an $S$-wave
behavior $\propto \beta$ as compared with the $P$--wave suppression
$\propto \beta^3$ for CP-even Higgs bosons
[$\beta=(1-4m_f^2/M_\Phi^2)^{1/2}$ is the velocity of the decay
fermions].  The QCD corrections to the partial decay width of the
CP-odd Higgs boson $A$ into heavy quark pairs are given in 
Ref.~\cite{drees,mathias1}, and for the charged Higgs particles in 
Ref.~\cite{hud}. \s 

Below the $t\bar t$ threshold, decays of the neutral Higgs bosons into
off-shell top quarks are sizeable, thus modifying the profile of the
Higgs particles significantly. Off-shell pseudoscalar branching ratios
reach a level of a few percent for masses above about 300 GeV for
small $\tb$ values. Similarly, below the $t\bar b$ threshold,
off-shell decays $H^+\to t^*\bar b \to b\bar b W^+$ are important,
reaching the percent level for charged Higgs boson masses above about
100 GeV for small $\tb$ values. The expressions for these  decays 
can be found in Ref.~\cite{1OFF}.

\subsection{Decays to gluons and electroweak gauge bosons} 

Since the $b$ quark couplings to the Higgs bosons may be strongly
enhanced and the $t$ quark couplings suppressed in the \MSSM, $b$
loops can contribute significantly to the Higgs-$gg$ couplings so that
the approximation $M_{Q}^{2} \gg M_{\Phi}^{2}$ cannot be applied any more
for $M_\Phi \lsim 150$~GeV, where this decay mode is important.
Nevertheless, it turns out {\it a posteriori} that this is an
excellent approximation for the QCD corrections in the range, where these
decay modes are relevant. For small $\tb$, the $t$ loop contribution 
is dominant and the decay width for $h,H\ra gg$ is given by eq.~(4)
with the appropriate factors for the $\Phi qq$ couplings; for a light 
pseudoscalar $A$ boson $\Gamma[A \ra gg]$ is also given by eq.~(4) with the 
change of the factor $95/4 \ra 97/4$. The
bottom and charm final states from gluon splitting can be added to 
the corresponding $b\bar b$ and $c\bar c$ decay modes, as in the SM case. \s

The decays of the neutral Higgs bosons to two photons and a photon
plus a $Z$ boson are mediated by $W$ and heavy fermion loops as in
the SM, and in addition by charged Higgs boson, sfermion and chargino 
loops\cite{HHG,gamma}; the partial decay widths can be found e.g. in   
Ref.~\cite{HHG} and are in general smaller than in the SM
except for the lightest $h$ boson in the decoupling limit
$M_A \sim M_H \sim M_{H^\pm} \gg M_Z$ since it is SM--like. 

The partial widths of the CP-even neutral Higgs bosons into $W$
and $Z$ boson pairs are obtained from the SM Higgs decay widths by
rescaling with the corresponding MSSM couplings. They are strongly
suppressed [due to kinematics in the case of $h$ and reduced couplings
for the heavy $H$], thus not playing a dominant role as in the SM.
Due to CP--invariance, the pseudoscalar $A$ boson does not
decay into massive gauge boson pairs at leading order. 

\subsection{Decays to Higgs and gauge boson pairs} 

The heavy CP-even Higgs particle can decay into light
scalar pairs as well as to pseudoscalar Higgs bosons pairs, $H \ra hh$ and
$H \ra AA$.  While the former is the dominant decay mode of $H$ in the
mass range $2M_h < M_H <2m_t$ for small values of $\tb$, the latter
mode occurs only in a marginal area of the MSSM parameter space. For
large values of $\tb$, these decays occur only if $M_A \sim M_h \lsim
M_H/2$, corresponding to the lower end of the heavy Higgs mass range,
and have branching ratios of 50\% each.  Since the $Hb\bar b$ Yukawa
coupling is strongly enhanced for large $\tb$, below threshold decays
$H \ra hh^*, AA^*$ with $A,h \ra b\bar{b}$ should also be included \cite{1OFF}.
The area of the parameter space in which the decay $h \ra AA$ is possible
\cite{sven} is ruled out by present data.  \s

The Higgs bosons can also decay into a gauge boson and a lighter Higgs
boson.  The branching ratios for the two body decays $A\ra hZ$ and
$H^+ \ra W^+h$ may be sizeable in specific regions of the MSSM
parameter space [small values of $\tb$ and below the $tt/tb$
thresholds for neutral/charged Higgs bosons]. 
Below-threshold decays into a Higgs particle and an off-shell gauge
boson turned out to be rather important in
the MSSM.  Off-shell $A \ra hZ^*$ decays are important for the
pseudoscalar Higgs boson for masses above about 130 GeV for small
$\tb$. The decay modes $H^\pm \to hW^*, AW^*$ reach branching ratios
of several tens of percent and lead to a significant reduction of the
dominant branching ratio into $\tau\nu$ final states to a level of
60\% to 70\% for small $\tb$. In addition, three-body $H \ra AZ^*$ and
$H\ra H^+W^{-*}$, which are kinematically forbidden at the two-body
level, can be sizeable for small $M_A$ values.  The 
expressions of the widths for these decay modes can be found in 
Ref.~\cite{1OFF}.

\subsection{Total Widths and Branching ratios} 

For large values of $\tb$ the decay pattern of the MSSM Higgs bosons is 
quite simple, a result of the strong
enhancement of the Higgs couplings  to down--type fermions. The neutral
Higgs bosons will decay into $b\bar{b}$ ($\sim 90\%$) and $\tau^+ \tau^-$ 
($\sim 10\%)$ pairs, and $H^\pm$ into $\tau \nu_\tau$ pairs below and $tb$ 
pairs above the top--bottom threshold. 
For the CP--even Higgs bosons $h(H)$, only when $M_h(M_H)$ 
approaches its maximal (minimal) value is this simple rule modified: 
in this decoupling 
limit, the $h$ boson is SM--like and decays into charm and gluons with a
rate similar to the one for $\tau^+ \tau^- $ [$\sim 5\%$] and in the high 
mass range, $M_h \sim 130$ GeV, into $W$ pairs with one of the $W$  bosons
being virtual; the $H$ boson will mainly decay into $hh$ and $AA$ final 
states. \s

For small values of $\tb\sim 1$ the decay pattern of the heavy neutral
Higgs bosons is much more complicated. The $b$ decays are in general
not dominant any more; instead, cascade decays to pairs of light Higgs
bosons and mixed pairs of Higgs and gauge bosons are important and
decays to $WW/ZZ$ pairs will play a role. For very large masses, they decay 
almost exclusively to top quark pairs. The decay pattern of  $H^\pm$ 
for small $\tb$ is similar to that at large $\tb$ except in 
the intermediate mass range where cascade decays to $Wh$ are dominant.
Off--shell three--body decays must be included and they 
provide a smooth transition from below to above threshold. The branching 
ratios  for $h,H,A$ and $H^\pm$ decays for $\tb=1.5$ are shown in Fig.2. \s

The total widths of the Higgs bosons are in general considerably
smaller than for the SM Higgs due to the absence or the suppression of 
the decays to $W/Z$ bosons which grow as $M_{H^0}^3$. 
The dominant decays for small $\tb$ are built-up by top quarks so that
the widths rise only linearly with $M_\Phi$. However, for large $\tb$ values, 
the decay widths scale in general like ${\rm tg}^2\beta$ and can 
become experimentally significant, for $\tb \gsim {\cal O}(30)$ and
for large $M_\Phi$. 

\begin{figure}[hbtp]
\vspace*{-2.3cm}
\hspace*{-3.5cm}
\begin{turn}{-90}%
\epsfxsize=12cm \epsfbox{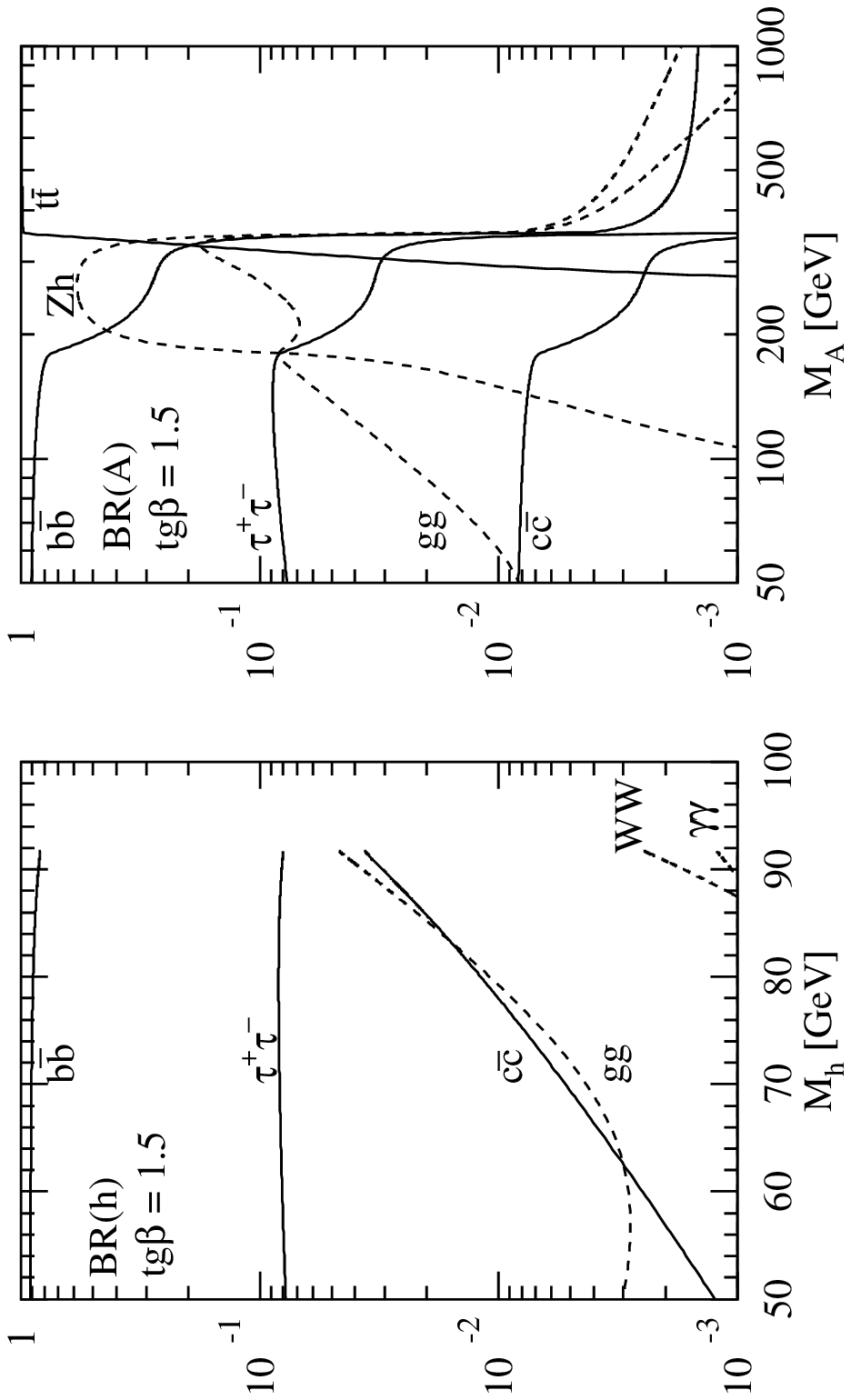}
\end{turn}
\vspace*{-4.2cm}
\end{figure}

\begin{figure}[hbtp]
\vspace*{-2.5cm}
\hspace*{-3.5cm}
\begin{turn}{-90}%
\epsfxsize=12.cm \epsfbox{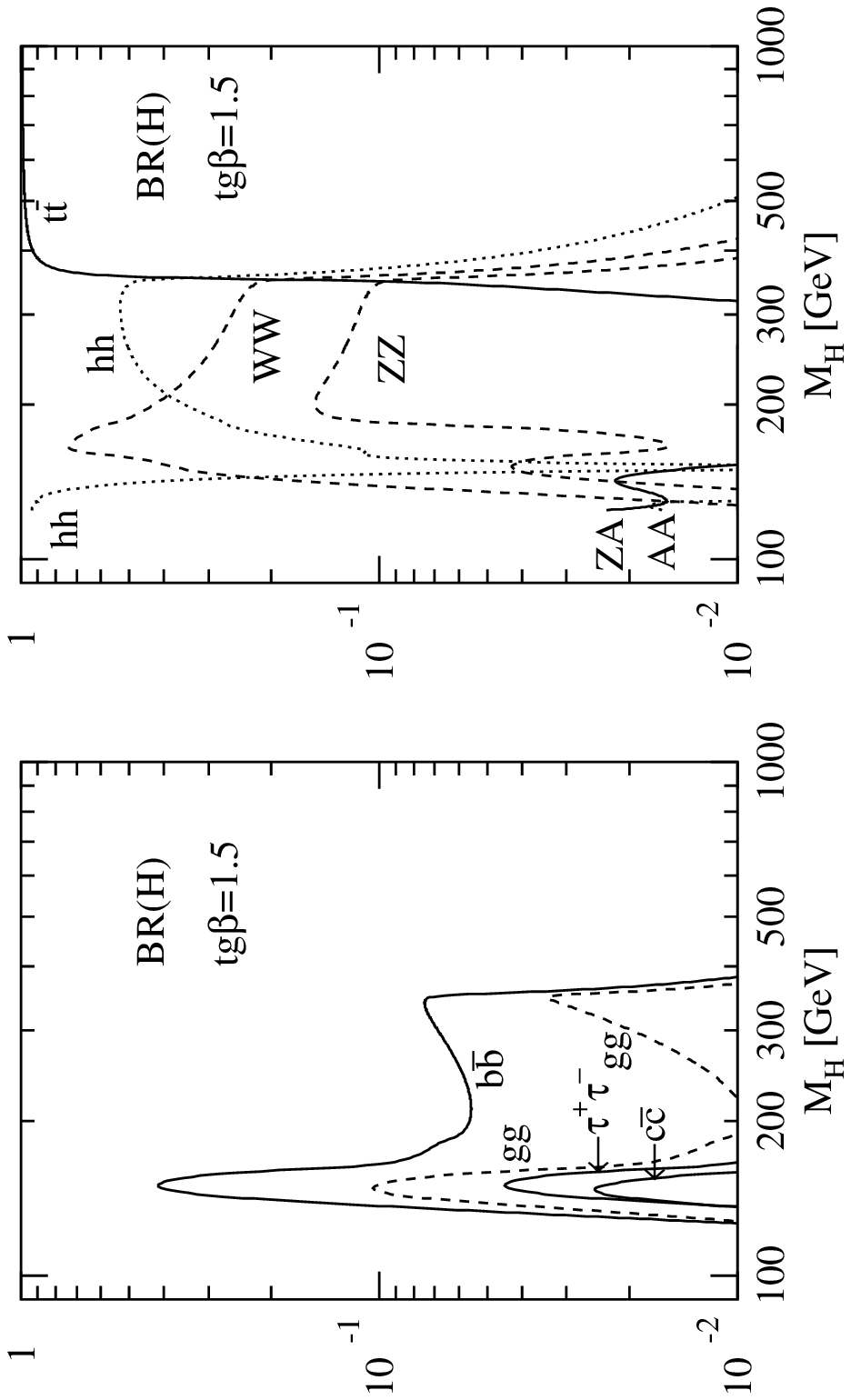}
\end{turn}
\vspace*{-3.6cm}

\vspace*{-1.7cm}
\hspace*{-3.5cm}
\begin{turn}{-90}%
\epsfxsize=12.cm \epsfbox{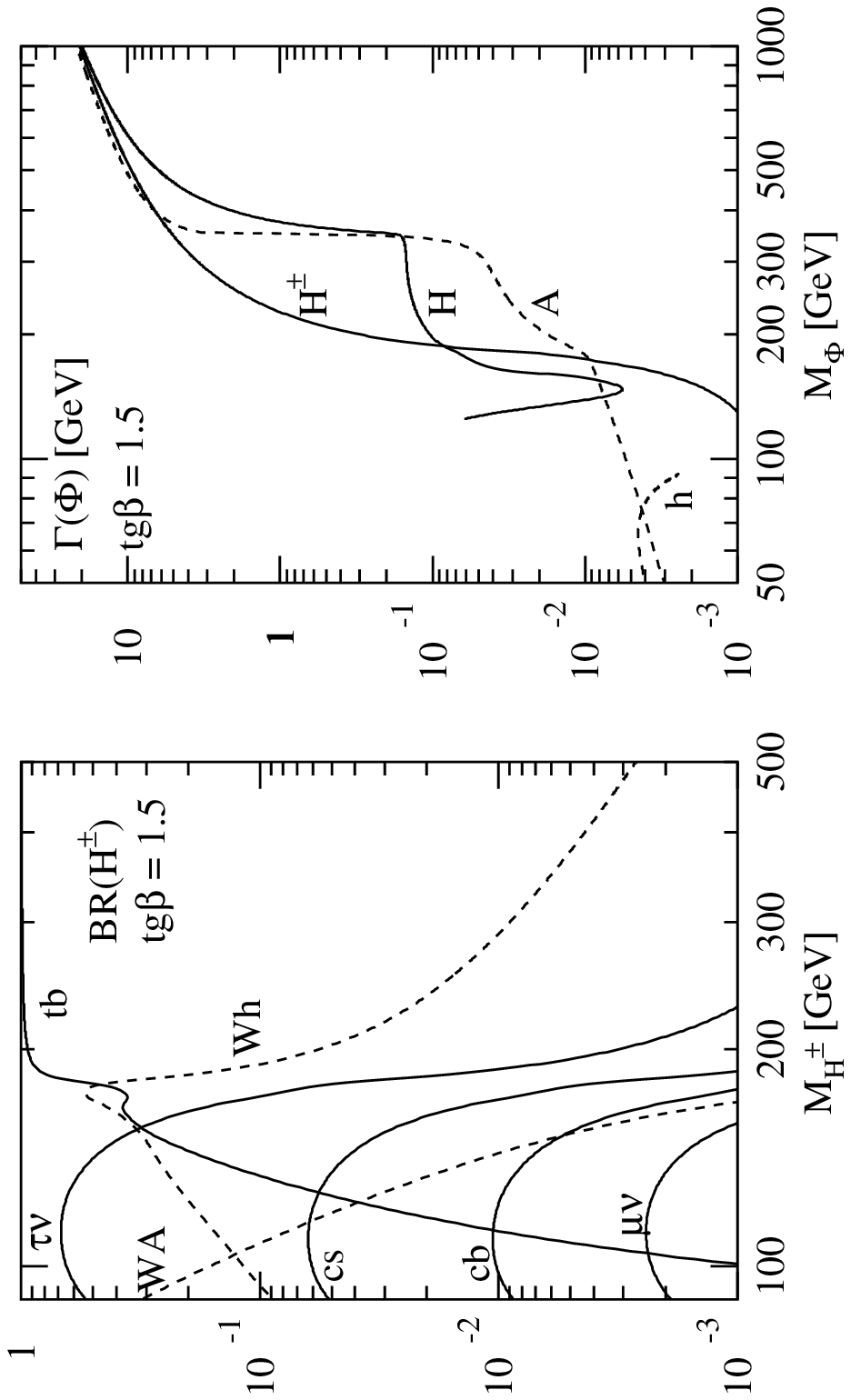}
\end{turn}
\vspace*{-3.1cm}

\caption[]{\label{fg:mssmbr} \it Branching ratios of the MSSM Higgs bosons
$h, A, H, H^\pm$ and their total decay widths $\Gamma(\Phi)$ as
functions of the Higgs mass $M_\Phi$ for $\tb=1.5$. The inputs in GeV are: 
$\mu=300, M_2=200, M_{\tilde{q}_L}=M_{\tilde{q}_R}=500$ and $A_t=1500$.}
\end{figure}

\section{Decays into Supersymmetric Particles} 

In the previous discussion, we have assumed that decay channels into
neutralinos, charginos and sfermions are shut. However, these channels
could play a significant role, since some of these particles 
[at least the lightest charginos, neutralinos and top squarks] 
can have masses in the ${\cal O}(100$ GeV) range or less. These 
decay modes will be discussed in this section. The partial widths of 
these decays can be found in  Refs.~\cite{R1,R2,R3}. 

\subsection{Decays into charginos and neutralinos}

Present experimental bounds on the SUSY particle masses, do not allow 
for SUSY decay modes of the lightest CP-even Higgs boson $h$ and of the
pseudoscalar Higgs boson $A$ for masses less than $\sim 100$ GeV, except 
for the decays into a pair of the lightest neutralinos.  However, 
whenever the $\chi_1^0\chi_1^0$ decay is kinematically allowed, the 
branching ratio is close to 100\% for positive $\mu$ values and small 
$\tb$ values. For $\mu<0$ the branching ratio never exceeds the 20\% level. 
The branching ratios become smaller for increasing $\tb$, except when $h$ 
reaches its maximal mass value since the $hb \bar{b}$ coupling is no longer 
enhanced. \s

For the heavier Higgs bosons $H,A$ and $H^\pm$, the branching ratios for the 
sum into all possible neutralino and chargino states are shown in Fig.~3.
Here mixing in the Higgs sector has been included for $\mu \neq 0$, and the
values $A_t=\sqrt{6}M_{\tilde{q}}$ [so-called ``maximal mixing"] and $A_b=0$, 
with $M_{\tilde{q}} =1$ TeV have been chosen. These branching ratios are 
always large except in three cases: $(i$) for $H$ in the mass range between 
140 and 200 GeV, especially if $\mu>0$, due to the large value of BR$(H \ra
hh)$; ($ii$) for small $A$ masses and negative $\mu$ values as discussed
above; and ($iii$) for $H^\pm$ just above the $t\bar{b}$ threshold if
not all the decay channels into the heavy $\chi$ states are open. \s

Even above the thresholds of decay channels including top quarks, the
branching ratios for the decays into charginos and neutralinos are
sizeable. For very large Higgs boson masses, they reach a common value
of $ \sim 40\%$ for $\tb =1.6$. In fact, as a consequence of the
unitarity of the diagonalizing $\chi$ mass matrices, the total widths of 
the three Higgs boson decays to charginos and neutralinos do not depend on 
$M_2$, $\mu$ or $\tb$ in the asymptotic regime $M_{\Phi} \gg m_\chi$,
giving rise to the branching ratio \cite{R2}
\begin{eqnarray}
{\rm BR}( \Phi \ra \sum_{i,j} \chi_i \chi_j) = \frac{ \left( 1+\frac{1}{3}
\tan^2 \theta_W \right) M_W^2 }{ \left( 1+\frac{1}{3} \tan^2
\theta_W \right) M_W^2 + m_t^2 \cot^2 \beta + m_b^2 \tan^2 \beta } \non 
\end{eqnarray}
Only the leading $t\bar{t}$, $b\bar{b}$ modes for neutral and the
$t\bar{b}$ modes for the charged Higgs bosons need to be included in the
total widths. This branching ratio is shown in Fig.~3 as a function of
$\tb$. It is always large, even for extreme values of $\tb \sim 1$ or $50$,
where it still is at the 20\% level. 
\bigskip

\begin{figure}[htbp]
\hspace*{0.3cm}
\centerline{\psfig{figure=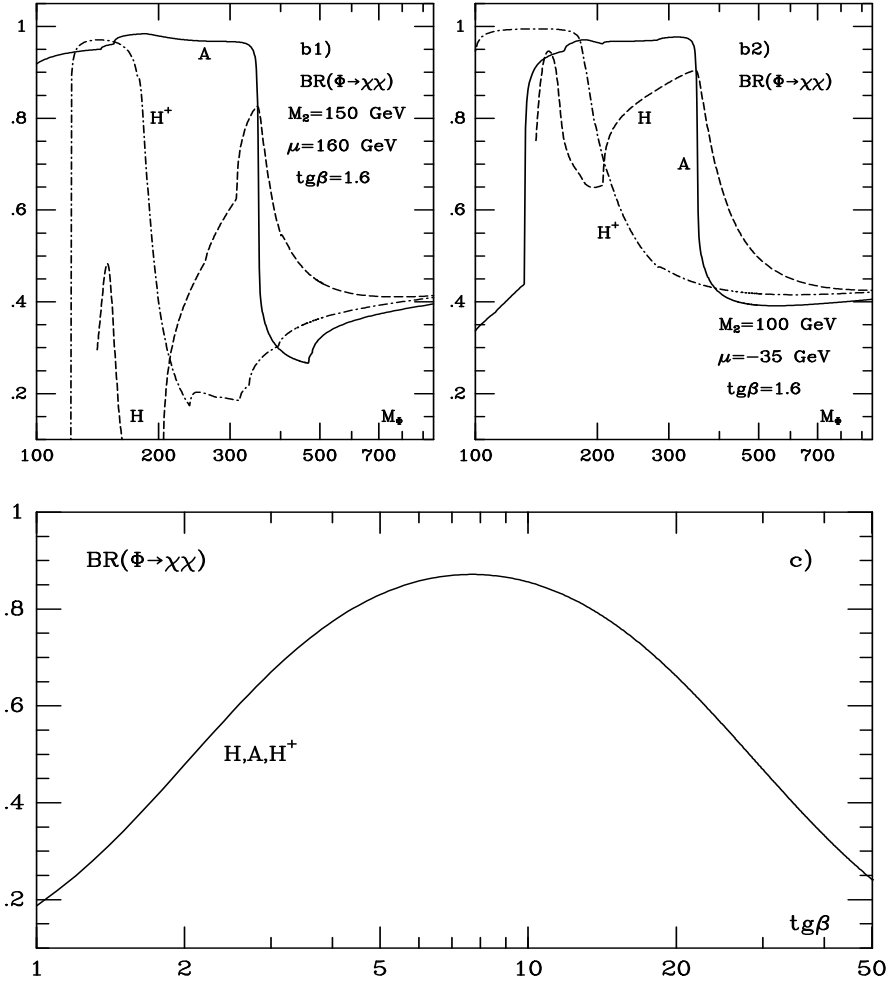,height=14.5cm,width=12.5cm}}
\vspace*{-2.4cm}
\caption[] { Up: the branching ratios of the decays of the heavy
$A$ [solid], $H$ [dashed] and $H^\pm$ [dot--dashed] Higgs bosons into the sum
of neutralino and chargino pairs as a function of the Higgs mass.
Down: the inclusive
$\chi \chi$ decay branching ratio as a function of $\tb$ in the
asymptotic region [$M_A \sim M_H \sim M_{H^\pm} = 1$ TeV $\gg m_{\chi}$];
From Ref.~\cite{R2}.}
\vspace{-4.5cm}
\end{figure}

\newpage

\subsection{Decays into Sfermions}

The decay widths of the heavy neutral CP--even and the charged Higgs bosons 
into first and second generation squarks and sleptons [the pseudoscalar $A$ 
boson cannot decay at tree-level into these states since the $A \tilde{f}_i 
\tilde{f}_i$ coupling is zero by virtue of CP--invariance and the $A 
\tilde{f}_1 \tilde{f}_2$ coupling is proportional to $m_f \sim 0$] 
are proportional to $ G_F M_W^4 /M_{\Phi}$
in the asymptotic regime $M_\Phi \gg m_{\tilde{f}}$. These decays are 
suppressed by the heavy Higgs mass and therefore unlikely to compete with the
dominant decay modes into top and/or bottom quarks [and to charginos and 
neutralinos] for which the decay widths grow as $M_\Phi$. \s

The situation is completely different for the decays into third generation
sfermions and in particular into top squarks \cite{R4}. Indeed, due to the 
large value of $m_t$ [which makes the mixing \cite{R5} in the stop sector 
possibly very large 
leading to a lightest top squark much lighter than the other squarks and 
even the top quark] the couplings of the Higgs bosons are strongly enhanced.
The partial widths up to mixing angle factors are proportional  to 
$G_F m_t^4/ (M_\Phi {\rm tg}^2\beta)$ and to $G_F m_t^2 (\mu+ 
A_t/\tb)^2 /M_\Phi$ where $A_t$ is the stop trilinear coupling. 
For small $\tb$ values and  not too heavy Higgs bosons, or for intermediate
values of $\tb$  and for $\mu$ and $A_t$ values of the order of $\sim 1$ TeV, 
the partial decay widths into top squarks can be very large and can compete 
with, and even dominate over, the decay channels into top quarks [and into
charginos/neutralinos]. Furthermore, decays into bottom squarks can also be 
important for large values of $\tb$ and $A_b$, since here also the mixing and
the couplings can be very large. \s

In order to have full control on these possibly dominant stop pair decays
of the Higgs bosons, QCD corrections must be included. They have been 
calculated recently \cite{R6} and found to be quite substantial, 
enhancing or suppressing the decay widths in Born approximation
by amounts up to 50\% and in some cases more. This is exemplified in Fig.~4, 
where the decay width for $H \ra \tilde{t}_1 \tilde{t}_1$ is shown for unmixed 
top squarks (up) and very large stop mixing (down). The decay widths are 
significantly larger for the case of mixing, 
being further increased by large QCD corrections up to nearly 50\%, whereas in 
the unmixed case the QCD corrections decrease the Born width significantly 
for the major part of the $\tilde{t}_1$ mass range; only close to the phase 
space boundary, the higher order contribution is positive. Large QCD 
corrections are also obtained for the decays $H, A \ra \tilde{t}_1 
\tilde{t}_2$ and $H^+ \ra \tilde{t} \tilde{b}$ as well as for the decay
$H \ra \tilde{b} \tilde{b}$. \s

The QCD corrections depend 
strongly on the gluino mass; however, for large gluino masses, the QCD 
correction is only logarithmically dependent on $m_{\tilde{g}}$. Contrary to 
the case of Higgs decays into light quarks, these QCD corrections cannot be 
absorbed into running squark masses since the latter are expected to be of 
the same order of magnitude as the Higgs boson masses. 

\begin{figure}[htbp]
\vspace*{-.8cm}
\centerline{\psfig{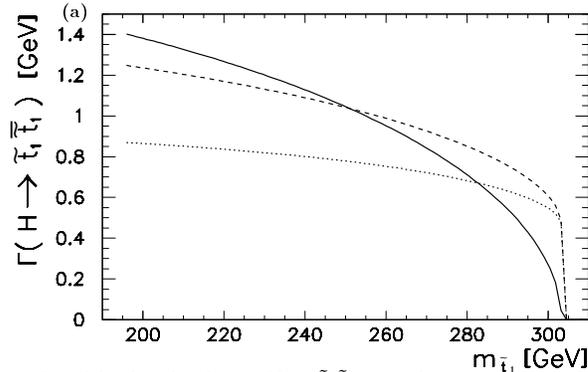}}
\vspace*{-11.cm}
\caption[] {Partial widths for the decay $H \ra \tilde{t}_1 
\tilde{t}_1$, as a function of $m_{\tilde{t}_1}$ with $M_H \sim 600$ GeV 
and $\tb=1.6$; $\mu = -300$ GeV, $A_t = -\mu \,\ctg\beta$ (up); $\mu = 
-300$ GeV, $A_t = 250$ GeV (down). The solid lines are for the 
Born approximation, while  the dashed and dotted lines are for the
widths including QCD corrections for $m_{\tilde{g}}=200$ GeV and 1 TeV 
respectively.}
\end{figure}

\vspace*{3.5cm}

\subsection{Decays in Minimal SUGRA} 

To discuss the SUSY decays, it is convenient to restrict oneself 
to the MSSM constrained by minimal Supergravity, in which the SUSY 
sector is described in terms of five universal parameters at the GUT 
scale: the common scalar mass $m_0$, the common gaugino mass $M_{1/2}$, 
the trilinear coupling $A$, the bilinear coupling $B$ and the higgsino 
mass $\mu$. These parameters evolve according to the RGEs, forming the 
supersymmetric particle spectrum at low energy.
The requirement of radiative electroweak symmetry breaking further
constrains the SUSY spectrum, since the minimization of the one--loop
Higgs potential specifies the parameter $\mu$ [to within a sign] and
also $B$. The unification of the $b$ and $\tau$ Yukawa couplings gives
another constraint: in the $\lambda_t$ fixed--point region, the value of
$\tb$ is fixed by the top quark mass through: $m_t \simeq (200~{\rm
GeV}) \sin\beta$, leading to $\tb \sim 1.5$. There also exists a
high--$\tb$ [$\lambda_b$ and $\lambda_\tau$ fixed--point] region for
which $\tb \sim$ 50.  If one also notes that moderate values of
the trilinear coupling $A$ have little effect on the resulting spectrum, 
then the whole SUSY spectrum will be a function of  $\tb$ which we
take to be $\tb=1.75$ and 50, the sign of $\mu$, $m_0$ which in practice we 
replace with $M_A$ taking the two illustrative values $M_A =300$ and 600 
GeV, and the common gaugino mass $M_{1/2}$ that are freely varied. \s

The decay widths of the heavy $H,A$ and $H^\pm$ Higgs bosons, 
into pairs of neutralinos and charginos [dashed lines], 
squarks [long--dashed  lines] and sleptons [dot--dashed lines], 
as well as the total [solid lines] and non--SUSY
[dotted--lines] decay widths, are shown in Fig.~5 for $\tb=1.75$, $\mu>0$
and two values of $M_A=300$ [left curves] and $600$ GeV [right
curves]. \s

For $M_A=300$ GeV, i.e. below the $t\bar{t}$ threshold, the widths of
the $H$ decays into inos and sfermions are much larger than
the non--SUSY decays. In particular, squark [in fact $\tilde{t}$ and 
$\tilde{b}$ only] decays are almost two--orders of magnitude larger when
kinematically allowed. The situation changes dramatically for larger $M_A$
when the $t\bar{t}$ channel opens up: only the decays into $\tilde{t}$ 
pairs when allowed are competitive with the dominant $H \ra t\bar{t}$
channel. Nevertheless, the decays into inos are still substantial having
BRs at the level of 20\%; the decays into sleptons never exceed a few percent.
\s

In the case of the pseudoscalar $A$, because of CP--invariance and the
fact that sfermion mixing is small except in the stop sector, only the
decays into inos and $A \ra \tilde{t}_1 \tilde{t}_2$
decays are allowed. For these channels, the situation is quite similar
to the case of $H$: below the $t\bar{t}$ threshold the decay width into
ino pairs is much larger than the non--SUSY decay widths [here
$\tilde{t}_2$ is too heavy for the $A \ra \tilde{t}_1 \tilde{t}_2$ decay
to be allowed], but above $2m_t$ only the $A \ra \tilde{t}_1
\tilde{t}_2$ channel competes with the $t\bar{t}$ decays. \s

For the charged Higgs boson $H^\pm$, only the decay  $H^+ \ra
\tilde{t}_1 \tilde{b}_1$ [when kinematically allowed] competes with the
dominant  $H^+\ra t\bar{b}$ mode, yet the $\tilde{\chi}^+ \tilde{
\chi}^0$ decays have a branching ratio of a few ten percent; the
decays into sleptons are at most of the order of one percent. \s

In the case where $\mu<0$, the situation is quite similar as above. For
large $\tb$ values, $\tb\sim 50$, all gauginos and sfermions are very
heavy and therefore kinematically inaccessible, except for the lightest
neutralino and the $\tau$ slepton. Moreover, the $b\bar{b}/\tau \tau$
and $t\bar{b}$/$\tau \nu$ [for the neutral and charged Higgs bosons
respectively] are enhanced so strongly, that they leave no chance for
the SUSY decay modes to be significant. Therefore, for large $\tb$, the
simple pattern of $bb/\tau\tau$ and $tb$ decays for heavy neutral and
charged Higgs bosons still holds true even when the SUSY decays are
allowed.

\begin{figure}[htbp]
\vspace*{.3cm}
\hspace*{0.6cm}
\centerline{\psfig{figure=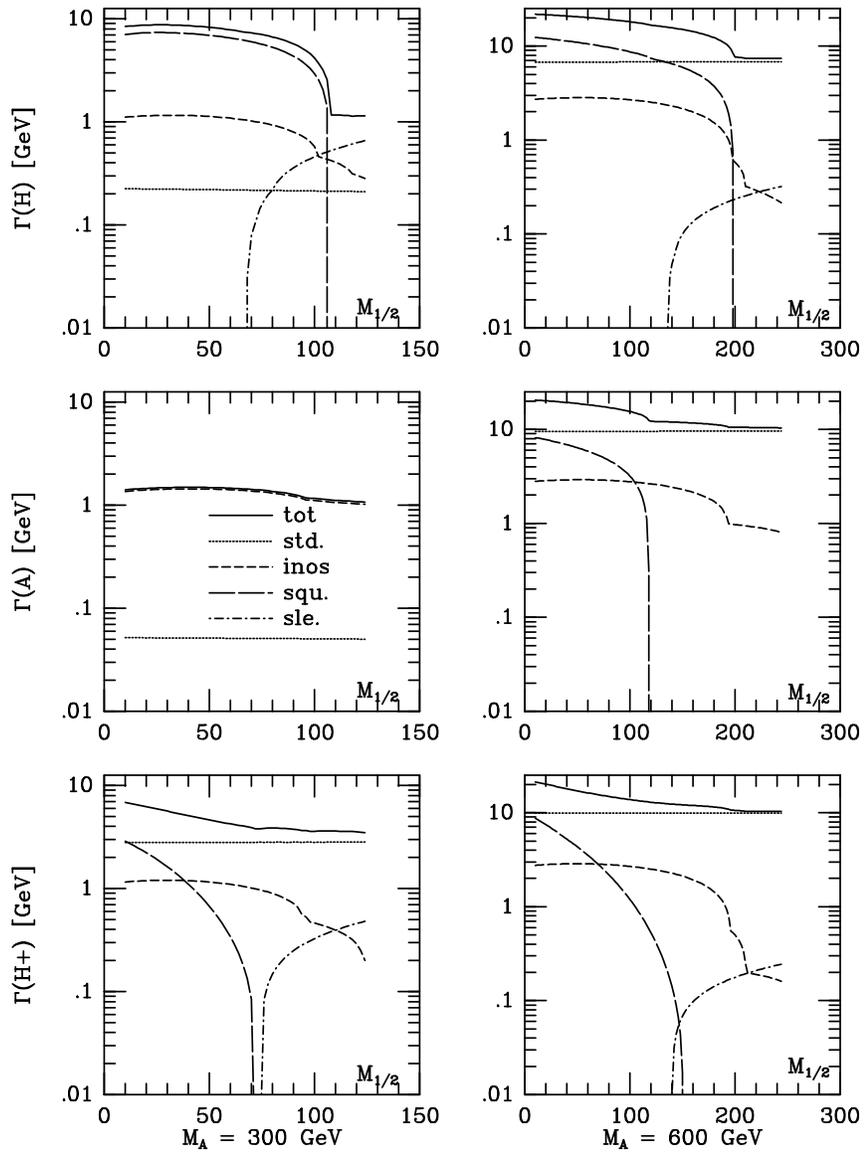,height=17cm}}
\vspace*{-1.9cm}
\caption[] {Decay widths for the SUSY channels of the
heavy CP--even, CP--odd and charged Higgs bosons, for $\tb=1.75$.
The total and the non--SUSY widths are also shown; From \cite{R3}.}
\end{figure}

\subsection{Decays into Light Gravitinos} 

Recently models \cite{R7} with a very light gravitino $\tilde{G}$, 
$m_{\tilde{G}} 
\leq 10^{-3}$ eV, have attracted some attention; see for instance 
Ref.~\cite{R8} and references therein. This interest 
was originally triggered by the resurgence of models of gauge mediated 
SUSY breaking and from the CDF $ee \gamma \gamma$ events. However, 
certain Supergravity models can also naturally accommodate a very
light gravitino \cite{R8}. \s

The couplings of the ``longitudinal'' (spin 1/2) components of the gravitino 
to ordinary matter are enhanced by the inverse of the $\tilde{G}$  mass
\cite{R7}; if 
$m_{\tilde{G}}$ is sufficiently small, this can compensate the suppression by 
the inverse Planck mass $M_P = 2.4 \cdot 10^{18}$ GeV that appears in all 
gravitational interactions. Since Gravitino couplings contain momenta of
the external particles, partial widths for decays into final states
containing (longitudinal) gravitinos depend very strongly on the mass of the
decaying particle. The neutral (charged) Higgs boson decay widths into a 
gravitino and neutralinos (charginos) are proportional to $M_\Phi^5$ 
and can be the dominant decay modes \cite{R9} for large values of $M_\Phi$. \s

This is shown in Fig.~6, where we plot the branching ratios of the $H,A$
and $H^\pm$ decays into light gravitinos and all possible combinations
of $\chi^0$ and $\chi^+$ as a function of $M_A$ and for a small value of 
$\tb=2$ and a gravitino mass of $10^{-4}$ eV. As can be seen, decays into 
light gravitinos could dominate the decays of all  three heavy Higgs 
bosons of the MSSM, if $M_A \geq 700$ GeV. For the lighter $h$ boson and 
for $A$ with $M_A \lsim$ 150 GeV the branching ratios cannot exceed 
a few percent for such a value of the $\tilde{G}$ mass. 

\begin{figure}[hbt]
\vspace*{-.2cm}
\centerline{\psfig{figure=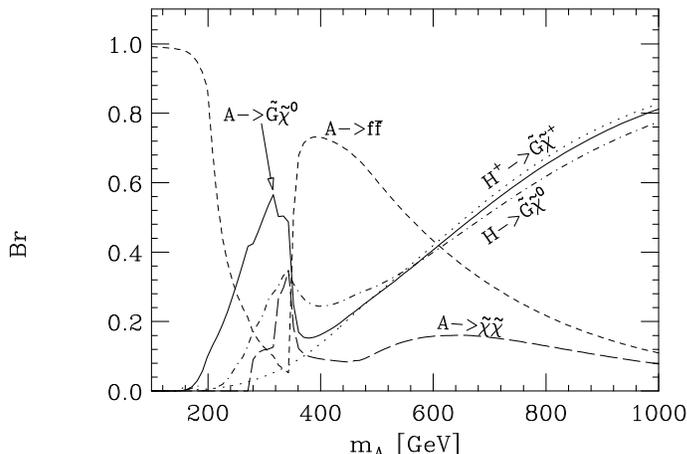,height=6cm,angle=-90}}
\vspace*{-.5cm}
\caption[]
{Branching ratios of the heavy Higgs boson decays into the sum
of charginos or neutralinos and a light $\tilde{G}$ as a function of 
$M_A$ for $M_2=300$ GeV, $\mu=-150$ GeV, $\tb=2$, $m_{\tilde{G}} 
=10^{-4}$ eV, $m_{\tilde{t}_L}= m_{\tilde{t}_R}=1$ TeV, $A_t =
\sqrt{6}$ TeV; From Ref.~\cite{R9}. } 
\end{figure}

\newpage

\section{SUSY Loop Effects} 

\subsection{SUSY--QCD corrections to the hadronic decays}

In the decays of the MSSM Higgs bosons into quark pairs, $\Phi \ra q\bar{q}$, 
besides the standard QCD corrections with virtual gluon exchange and gluon 
emission in the final state, one needs to include the contributions of the 
partner squark and gluino exchange diagrams. These SUSY--QCD corrections 
have been calculated by several authors \cite{QCDqq,sola} and found to be
rather substantial for not too heavy squark and gluino masses. For the 
electroweak corrections, see Ref.~\cite{SUSYEW}. \s

In the case of the  $h, A, H \ra b\bar{b}$ decays, the SUSY-QCD 
corrections can be very large reaching the level of several ten
percent for moderate values of $m_{\tilde{b}}$ and $m_{\tilde{g}}$;
in particular corrections of the order of 50 to 60 \% can be obtained 
for large values of $\tb$ if $m_{\tilde{g}} \sim 200$ GeV. 
In general, the sign of the correction is opposite to the sign of $\mu$.
The corrections relative to the Born terms are shown in Fig.~7 as a function 
of the $\tilde{b}$ mass for several values of $\tb$ and $M_A=60$ GeV.   
As can be seen the corrections decrease with increasing $m_{\tilde{b}}$,
but they can still be at the level of a few ten percent for  
$m_{\tilde{b}} \sim $ a few hundred GeV. The situation is
similar for the asymptotic behavior with $m_{\tilde{g}}$ 
as it takes a long time for the gluino to decouple: for $m_{\tilde{g}} 
\sim 1$ TeV, one is still left with substantial QCD corrections for not 
too heavy bottom squarks.  

\begin{figure}[htbp]
\vspace*{-2cm}
\centerline{\psfig{figure=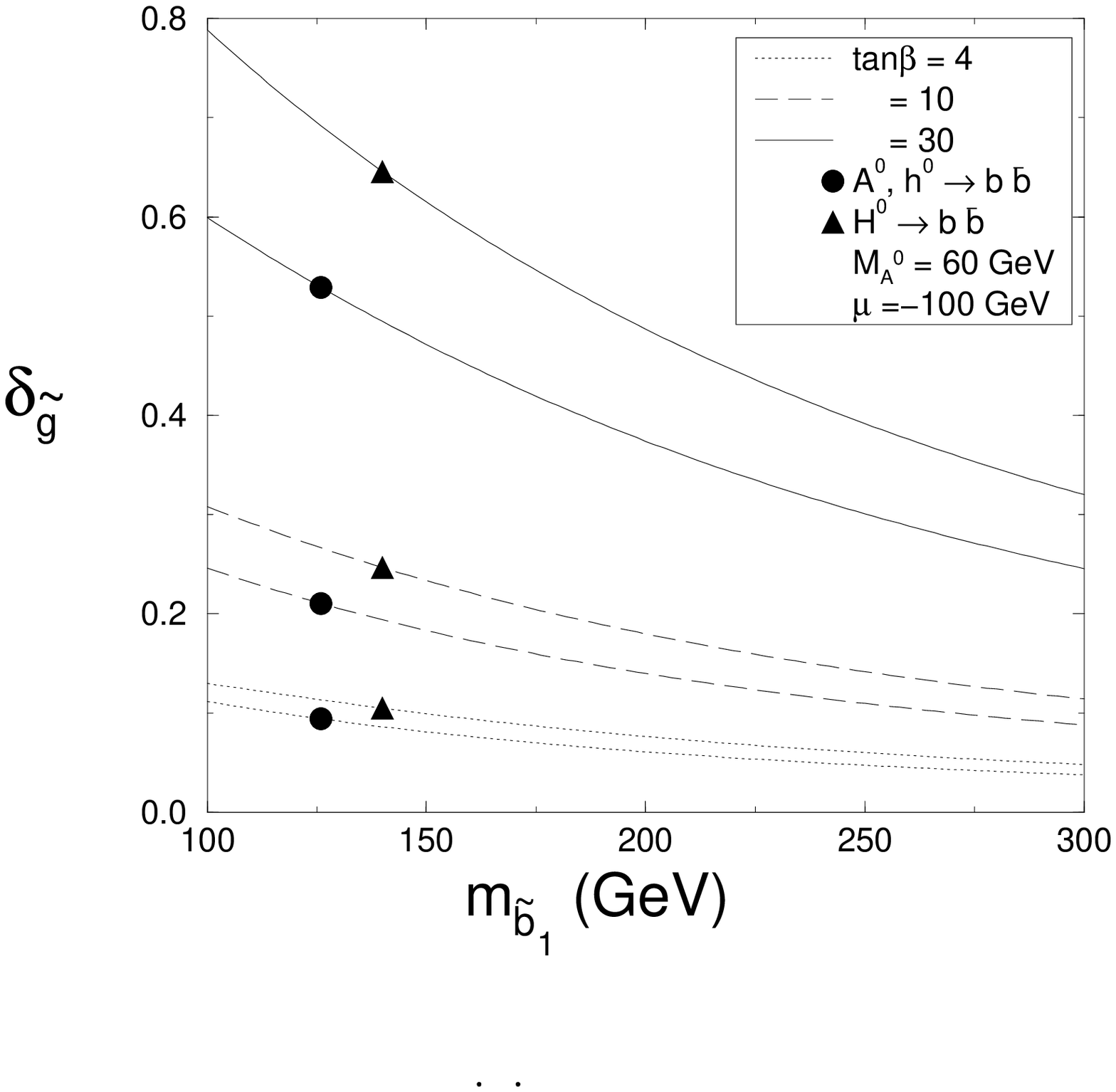,height=14cm}}
\vspace*{-5.6cm}
\caption[] {
The sbottom--gluino QCD correction to the decays $h,H,A \ra b\bar{b}$ 
normalized to the Born widths as a function of $m_{\tilde{b}_1}$ for 
various values of $\tb$ and fixed values of $\mu$ and $M_A$; taken from 
Ref.~\cite{sola}.}
\end{figure}

\newpage 

For heavier Higgs bosons, the SUSY--QCD corrections to the decays $H,A \ra 
t\bar{t}$ and $H^+ \ra t\bar{b}$ can also be large \cite{QCDqq,sola},
reaching the level of several 10\%. \s

In the gluonic decay modes $h,H \ra gg$, squark and in particular top squark 
loops  must be included [squark loops do not contribute to the $Agg$ coupling 
because of CP--invariance] since these contributions are significant for squark 
masses $M_{\tilde{Q}} \lsim 500$ GeV and small $\tb$ values. This can be seen 
in Fig.~8 where the ratio of the gluonic decay width of the $h$ boson with 
and without the squark contributions is shown as a function $M_{\tilde{Q}}$
for $\tb=1.5,30$. The QCD corrections \cite{SQCD} to the squark contribution 
have been calculated in the heavy squark mass limit, and are approximately of 
the same size as the the QCD corrections to the top quark contribution. 
A reasonable approximation [within about 10\% ] to the
gluonic decay width can be obtained by multiplying the full lowest order
expression [including quark and squark contributions] with the relative QCD 
corrections including only quark loops. \s

Note that the QCD correction to the squark contribution to the $h \ra \gamma 
\gamma$ coupling, which will be discussed later, has also been calculated 
\cite{hgaga}: in the heavy squark mass limit and relatively to the Born term,
the correction is $8\alpha_s/3\pi$ [compared to $-\alpha_s/\pi$ for the top
quark loop] and is therefore small.  

\begin{figure}[hbt]
\vspace*{-1.4cm}
\hspace*{1.1cm}
\epsfxsize=8cm \epsfbox{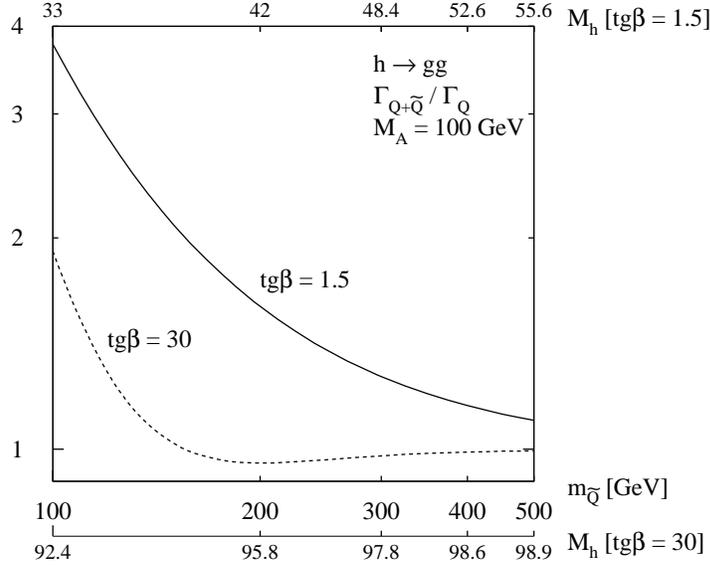}
\vspace*{-2.7cm}
\caption[]{Ratio of the QCD-corrected decay width
$\Gamma(h \ra gg)$ with and without squark loops for two values of $\tb = 1.5,
30$ as a function of the common squark mass $M_{\tilde{Q}}$. $M_A=100$ GeV
and the second axes show the corresponding values of $M_h$; from 
Ref.\cite{habil}.}
\end{figure}

\newpage

\subsection{SUSY Loop Effects in $h \ra \gamma \gamma$}

In the decoupling limit, $M_H \sim M_A \sim M_{H^+} \gg M_Z$, the lightest 
SUSY Higgs boson $h$ has almost 
the same properties as the SM Higgs particle $H^0$ and the MSSM and
SM Higgs sectors look practically the same. In the case where no genuine 
SUSY particle and no additional Higgs boson have been found at future 
machines, the task of discriminating between the lightest SUSY and the 
SM Higgs boson is challenging. A way to discriminate between the two
in this decoupling regime is to look at loop induced Higgs  boson 
couplings such as the couplings to $gg$, $Z\gamma$ and $\gamma \gamma$.
In the SM, these couplings are mediated by heavy quark and $W$ 
boson loops [only quark loops for the $H^0gg$ coupling]. In supersymmetric 
theories, additional contributions will be induced by loops with charged 
Higgs bosons, charginos and sfermions. \s

The $hgg$ coupling, which can be measured in the decays $h \ra gg$ or at 
the LHC in the dominant production mechanism $gg \ra h$, has been discussed 
previously. The $hZ\gamma$ coupling, which could be measured for $M_h 
<M_Z$ in the decay $Z\ra h \gamma$, at a high--luminosity $e^+e^-$ collider 
running at the Z--peak, or in the reverse decay $h \ra Z\gamma$ 
if $M_h >M_Z$ at the LHC, has been discussed in Ref.~\cite{hgaZ}:
the SUSY--loop effects are large only in extreme situations, and are 
unlikely to be seen in these decays. We will discuss here only the $h 
\gamma \gamma$ coupling \cite{hgaga} which could be measured in the decays
$h \ra \gamma \gamma$ with the Higgs boson produced at LHC in the 
$gg \ra h$ mechanism or at future high--energy and high--luminosity 
$\ee$ colliders in the process $\ee \ra h \nu \bar{\nu}$, and most promising
in the s--channel single Higgs production in the fusion process $\gamma \gamma 
\ra h$, with the photons generated by Compton--back scattering of laser light
[a measurement with a precision of the order of 10\% could be feasible
in this case]. \s

The contributions of charged Higgs bosons, sleptons and the scalar
partners of the light quarks including the bottom squarks are extremely
small. This is due to the fact that these particles do not couple to the
Higgs boson proportionally to the mass, and the amplitude is damped by
inverse powers of the heavy mass squared; in addition, the couplings are
small and the amplitude for spin--0 particles is much smaller than the dominant
$W$ amplitude. \s

The contribution of the charginos to the two--photon decay width
can exceed the 10\% level for masses close to $m_\chi \sim 100$
GeV, but it becomes smaller with higher masses. The deviation of the
$\Gamma(h \ra \gamma \gamma)$ width from the SM value induced by 
charginos with masses $m_\chi=250$ and $400$ GeV is shown in 
Fig.~9, as a function of $M_2$ [$\mu$ is fixed by $m_\chi$] for 
$\tb=1.6$ and $50$. For chargino masses above $m_\chi \gsim 250$ 
GeV [i.e. slightly above the limit where charginos can be produced 
at e.g. a 500 GeV $\ee$ collider], the deviation is less than $ \sim 8\%$ 
for the entire SUSY parameter space. The deviation drops by a factor of 
two if the chargino mass is increased to 400 GeV. \s

Because its coupling to the lightest Higgs boson can be strongly
enhanced, the top squark can generate sizeable contributions to the
two--photon decay width of the $h$ boson. 
For stop masses in the $\sim 100$ GeV range, the
contribution could reach the level of the dominant $W$ boson
contribution and the interference is constructive increasing drastically
the decay width. For $\tilde{t}_1$ masses around 250 GeV, the deviation of 
the $h \ra \gamma \gamma$ decay width from the SM value can be still at 
the level of 10\% for a very large off--diagonal entry in the stop mass 
matrix, $m^{LR}_t \gsim 1$ TeV; Fig.~9. For larger masses, the deviation 
drops $\sim 1/m_{\tilde{t}_1}^2$ and the effect on the decay width is below
$2\%$ for $m_{\tilde{t}_1} \sim 400$ GeV even at $m^{LR}_t \sim 1$ TeV. 
For small values of $m^{LR}_t$, the deviation does not exceed $-8\%$
even for a light top squark $m_{\tilde{t}_1} \sim 250$ GeV.

\begin{figure}[htbp]
\vspace*{-2cm}
\centerline{\psfig{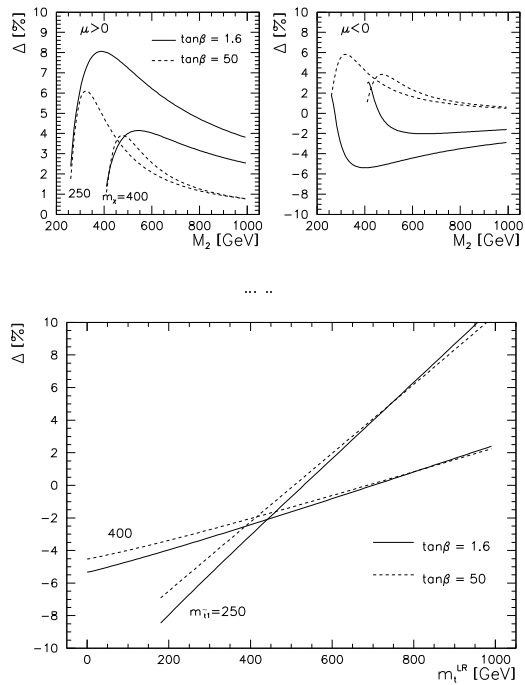}}
\vspace*{-2.cm}
\caption[] {
The deviations of the SUSY Higgs coupling to two photons from the 
SM value [in \%] for two values of $\tb=1.6$ and 50
and the loops masses $m_i =250$ and $400$ GeV. Deviations 
due to the chargino loops as a function of $M_2$ for both signs of $\mu$
(up), and deviations due to the top squark loops (down) as a function of 
$m_{t}^{LR}$; from Ref.~\cite{hgaga}. }
\end{figure}

\section{The program HDECAY}

Finally, let me make some propaganda and shortly describe the fortran 
code HDECAY \cite{hdecay}, 
which calculates the various decay widths and the branching ratios of 
Higgs bosons in the SM and the MSSM and which includes: 

(a) All decay channels that are kinematically allowed and which have
branching ratios larger than $10^{-4}$, {\it y compris} the loop
mediated, the three body decay modes and in the MSSM the cascade and the
supersymmetric decay channels.

(b) All relevant two-loop QCD corrections to the decays into quark pairs and 
to the quark loop mediated decays into gluons are incorporated in the most 
complete form; the small leading electroweak corrections are also included. 

(c) Double off--shell decays of the CP--even Higgs bosons into massive
gauge bosons which then decay into four massless fermions, and all 
all important below--threshold three--body decays discussed previously. 

(d) In the MSSM, the complete radiative corrections in the effective
potential approach with full mixing in the stop/sbottom sectors;
it uses the renormalisation group improved values of the Higgs masses
and couplings and the relevant leading next--to--leading--order corrections 
are also implemented. 

(e) In the MSSM, all the decays into SUSY particles (neutralinos, charginos,
sleptons and squarks including mixing in the stop, sbottom and stau 
sectors) when they are kinematically allowed. The SUSY particles are also 
included in the loop mediated $\gamma \gamma$ and $gg$ decay channels.

\smallskip

The basic input parameters, fermion and gauge boson masses and total
widths, coupling constants and in the MSSM, soft--SUSY breaking
parameters can be chosen from an input file. In this file several flags
allow to switch on/off or change some options [{\it e.g.} chose a
particular Higgs boson, include/exclude the multi--body or SUSY decays,
or include/exclude specific higher--order QCD corrections]. The results for the
many decay branching ratios and the total decay widths are written to
several output files with headers indicating the processes and giving
the input parameters. \s

The program is written in FORTRAN and has been tested on several
machines: VAX stations under the operating system VMS and work
stations running under UNIX. All the necessary subroutines [e.g. for
integration] are included. The program is lengthy [more than 5000
FORTRAN lines] but rather fast, especially if some options [as decays
into double off-shell gauge bosons] are switched off. 

\section{Summary}
In this talk, the decay modes of the Standard and Supersymmetric Higgs bosons
in the MSSM, have been reviewed and updated. The relevant higher--order 
corrections which are dominated by the QCD radiative corrections and the
off--shell [three--body] decays have been discussed. In the MSSM, the SUSY 
decay modes,
and in particular the decays into charginos, neutralinos, and top squarks [as
well as decays into light gravitinos] can be very important in large regions
of the parameter space. The SUSY--loop contributions to the standard
decays into quarks, gluons and photons of the MSSM Higgs bosons can also
be important for not too heavy SUSY particles. 
The total decays widths of the Higgs bosons and the various branching ratios 
in the SM and in the MSSM, including the previous points can be obtained using 
the program HDECAY.

\section*{Acknowledgments}
I would like to thank Joan Sol\`a and the Organizing Committee for the 
invitation to this Workshop and for their efforts to make the meeting 
very fruitful.


\section*{References}

\begin{thebibliography}{99}

\bibitem{HHG} For a review on the Higgs sector of the SM and the MSSM ,
see J.F. Gunion, H.E. Haber, G.L. Kane and S. Dawson, {\it The Higgs 
Hunters Guide}, Addison-Wesley, Reading 1990.

\bibitem{REV} For a review on Higgs physics at future hadron and $\ee$ 
colliders see e.g., A.\ Djouadi, Int.\ J.\ Mod.\ Phys.\ {\bf A10} (1995) 1.

\bibitem{LEPbound} P. Janot, EuroConference on High--Energy Physics, 
Jerusalem, 1997. 

\bibitem{DATA} For a recent account on the constraints on the SM and MSSM 
Higgs masses, see M. Carena, P.M. Zerwas et al., {\it Higgs Physics at
LEP2}, CERN 96--01, G. Altarelli, T. Sj\"ostrand and F. Zwirner 
(eds.). 

\bibitem{RC}Y.\ Okada, M.\ Yamaguchi and T.\ Yanagida, Prog.\ Theor.\ Phys.\ 
{\bf 85} (1991) 1; H.\ Haber and R.\ Hempfling, Phys.\ Rev.\ Lett.\ {\bf 66} 
(1991) 1815; J.\ Ellis, G.\ Ridolfi and F.\ Zwirner, Phys.\ Lett.\ {\bf B257} 
(1991) 83; R.\ Barbieri, F.\ Caravaglios and M.\ Frigeni, Phys.\ Lett.\ {\bf 
B258} (1991) 167; A. Hoang and R. Hempfling, Phys. Lett. B331 (1994) 99; J. 
Casas, J.\ Espinosa, M.\ Quiros and A. Riotto, Nucl. Phys. {\bf B436} (1995) 3; 
M.\ Carena, J.\ Espinosa, M.\ Quiros and C.E.M.\ Wagner, Phys.\ Lett.\ {\bf 
B355} (1995) 209. 
\bibitem{quiros} See the talk given by M. Quiros, these proceedings. 
\bibitem{habil} M.\ Spira, Report CERN-TH/97-68, hep-ph/9705337.
\bibitem{ho} See the talk given by W. Hollik at this Workshop, hep-ph/9711489, 
to appear in the proceedings. 
\bibitem{QCD} For an update of the effect of QCD corrections
to the hadronic decay widths, see A.\ Djouadi, M.\ Spira and 
P.M.\ Zerwas, Z.\ Phys.\ {\bf C70} (1996) 427; see also J. Kamoshita, Y.
Okada and M. Tanaka, Phys. Lett. {\bf B391} (1997) 124; and
Z. Kunszt, S. Moretti and W.J. Stirling Z. Phys. {\bf C74} (1997) 479. 

\bibitem{drees} E.\ Braaten and J.P.\ Leveille, Phys.\ Rev.\ {\bf D22} (1980)
715; M.\ Drees and K.\ Hikasa, Phys.\ Lett.\ {\bf B240} (1990) 455; (E) {\bf 
B262} (1991) 497.

\bibitem{russ} S.G.\ Gorishny, A.L.\ Kataev, S.A.\ Larin and L.R.\ Surguladze,
Mod.\ Phys.\ Lett.\ {\bf A5} (1990) 2703; Phys.~Rev.~{\bf D43} (1991) 1633; 
A.L.~Kataev and V.T.~Kim, Mod.~Phys.~Lett.~{\bf A9} (1994) 1309; 
L.R.~Surguladze, Phys. Lett. {\bf 341} (1994) 61; K.G.~Chetyrkin, Phys.\ 
Lett.\ {\bf B390} (1997) 309.

\bibitem{broad} N.~Gray, D.J.~Broadhurst, W.~Grafe and K.~Schilcher,
Z.~Phys.~{\bf C48} (1990) 673.

\bibitem{narison} S.~Narison, Phys.~Lett.~{\bf B341} (1994) 73.

\bibitem{runmass} S.G.\ Gorishny, A.L.\ Kataev, S.A.\ Larin and L.R.\
Surguladze, Mod.\ Phys.\ Lett.\ {\bf A5} (1990) 2703; Phys.~Rev.~{\bf D43}
(1991) 1633

\bibitem{mathias1} R. Harlander and M. Steinhauser, Phys. Rev. {\bf D56} (1997)
3980. 

\bibitem{1OFF} The discussions on the three-body decays are based on: 
A.\ Djouadi, J.\ Kalinowski and P.M.\ Zerwas, Z.\ Phys.\ {\bf C70} (1996) 437;
see also S. Moretti and W.J. Stirling, Phys. Lett. {\bf B347} (1995) 
291; (E) {\bf B366} (1996) 451.

\bibitem{rcelw} For a summary and a complete set of references, see 
B.A. Kniehl, Phys, Rep. {\bf 240} (1994) 211.

\bibitem{hgglo} J.\ Ellis, M.K.\ Gaillard and D.V.\ Nanopoulos, Nucl.\ Phys.\
{\bf B106} (1976) 292; A.I.\ Vainshtein, M.B.\ Voloshin, V.I.\ Sakharov 
and M.A.\ Shifman, Sov.\ J.\ Nucl.\ Phys.\ {\bf 30} (1979) 711.

\bibitem{higgsqcd} M.\ Spira, A.\ Djouadi, D.\ Graudenz and P.M.\ Zerwas,
Nucl.\ Phys.\ {\bf B453} (1995) 17.

\bibitem{hgg} T.~Inami, T.~Kubota and Y.~Okada, Z.~Phys.~{\bf C18} (1983) 69; 
A.\ Djouadi, M.\ Spira and P.M.\ Zerwas, Phys.\ Lett.\ {\bf B264} (1991) 440.  

\bibitem{mathias2} K.G. Chetyrkin, B.A. Kniehl and  M. Steinhauser,
Phys. Rev. Lett. {\bf 79} (1997) 353. 

\bibitem{hvv0} B.W.~Lee, C.~Quigg and H.B.~Thacker, Phys.~Rev.~{\bf D16} (1977)
1519.
\bibitem{hvvlam} A.\ Ghinculov, Nucl.\ Phys.\ {\bf B455} (1995) 21; 
A.\ Frink, B.\ Kniehl, D.\ Kreimer, and K.\ Riesselmann, Phys.\ Rev.\ {\bf D54}
(1996) 4548.
\bibitem{hvv} T.G.~Rizzo, Phys.~Rev.~{\bf D22} (1980) 389; 
W.-Y.~Keung and W.J.~Marciano, Phys.~Rev.~{\bf D30} (1984) 248.
\bibitem{2OFF} See e.g., R.N.\ Cahn, Rep.\ Prog.\ Phys.\ {\bf 52} (1989) 389. 
\bibitem{hud} A. Mendez and A. Pomarol, Phys. Lett. {\bf B252} (1990) 461; 
C.S. Li and R.J. Oakes, Phys. Rev. {\bf D43} (1991) 855; 
A.~Djouadi and P.~Gambino, Phys.~Rev.~{\bf D51} (1995) 218.
\bibitem{gamma} J.F. Gunion and H. Haber,   Phys. Rev. {\bf D48} (1993) 5109;
G. L. Kane, G. D. Kribs, S. P. Martin and J. D. Wells, Phys. Rev. {\bf D53} 
(1996) 213;  B. Kileng, P. Osland and P.N. Pandita, Z. Phys. {\bf C71}
(1996) 87. 
\bibitem{sven} S. Heinemeyer and W. Hollik, Nucl. Phys. {\bf B474} (1996) 32.
\bibitem{R1} J.F. Gunion and H.E. Haber, Nucl.\ Phys.\ {\bf B272} (1986) 1;
{\bf B278} (1986) 449; {\bf B307} (1988) 445; (E) hep-ph/9301201.
\bibitem{R2} A.\ Djouadi, P. Janot, J.\ Kalinowski and P.M.\ Zerwas, 
Phys. Lett. {\bf B376} (1996) 220;  A.\ Djouadi, J.\ Kalinowski and 
P.M.\ Zerwas, Z. Phys. {\bf C57} (1993) 569. 
\bibitem{R3} A.\ Djouadi, J.\ Kalinowski, P.\ Ohmann and P.M.\ Zerwas, 
Z.\ Phys.\ {\bf C74} (1997) 93.
\bibitem{R4} J.F. Gunion and H.E. Haber, Phys.\ Rev.\ {\bf D37} (1988) 2515; 
A. Bartl et al., Phys. Lett. {\bf B373} (1996) 117.
\bibitem{R5} J. Ellis and D. Rudaz, Phys. Lett. {\bf B128} (1983) 248; 
K. Hikasa and M. Drees, Phys. Lett. {\bf B252} (1990) 127. 
\bibitem{R6} A.\ Bartl, H.\ Eberl, K.\ Hidaka, T.\ Kon, W.\ Majerotto
and Y.\ Yamada, Report UWThPh-1997-03, hep-ph/9701398; A.\ Arhrib, 
A.\ Djouadi, W.\ Hollik and C.\ J\"unger, Report KA-TP-30-96, hep-ph/9702426;
see also the talks of A. Bartl and W.\ Majerotto in these proceedings.  
\bibitem{R7} P. Fayet, Phys. Lett. {\bf 70B}, 461 (1977) and {\bf B175}, 
471 (1986). 
\bibitem{R8} S. Ambrosiano, G.L. Kane, G.D. Kribs, S.P. Martin and S. Mrenna,
Phys. Rev. {\bf D54}, 5395 (1996); J. Ellis, J.L. Lopez and D.V. Nanopoulos, 
Phys. Lett. {\bf B394} (1997) 354. 
\bibitem{R9} A. Djouadi and M. Drees, Phys. Lett. {\bf B407} (1997) 243. 
\bibitem{QCDqq} C.S. Li and J.M. Yang, Phys. Lett. {\bf B315} (1993) 367; 
A. Dabelstein, Nucl. Phys. {\bf B456} (1995) 25; A. Bartl, H. Eberl, 
K. Hidaka, T. Kon, W. Majerotto and Y. Yamada Phys. Lett. {\bf B378} (1996) 
167; R.A. Jimenez and J. Sola, Phys. Lett. {\bf B389} (1996) 53; see also 
the talk of W.\ Majerotto in these proceedings.
\bibitem{sola} J.A. Coarasa, R.A. Jimenez and J. Sol\`a, Phys. Lett. 
{\bf B389} (1996) 312. 
\bibitem{SUSYEW} A. Dabelstein and W. Hollik, Report MPI-PH-93-86; 
see also the talk by W. Hollik, these proceedings. 
\bibitem{SQCD} S.\ Dawson, A.\ Djouadi and M.\ Spira, Phys.\ Rev.\ Lett.\
{\bf 77} (1996) 16. 
\bibitem{hgaga} A. Djouadi, V. Driesen, W. Hollik and J.I. Illana,
hep-ph/9612362.
\bibitem{hgaZ} A. Djouadi, V. Driesen, W. Hollik and A. Kraft, 
hep-ph/9701342. 
\bibitem{hdecay} A. Djouadi, J. Kalinowski and M. Spira, 
hep-ph/9704448; the program can be obtained from the net at
http://www.lpm.univ-montp2.fr/$\tilde{}$djouadi/program.html or
http://wwwcn.cern.ch/$\tilde{}$mspira. 


\end{thebibliography}
\end{document}